\pdfoutput=1 

\documentclass[preprints,article,accept,moreauthors,unicode]{Definitions/mdpi} 
\firstpage{1} 
\pubvolume{1}
\issuenum{1}
\articlenumber{0}
\pubyear{2026}
\copyrightyear{2026}
\datereceived{28 May 2026} 
\daterevised{23 August 2026} 
\dateaccepted{30 August 2026 } 
\datepublished{ } 

\Title{Neutrino Astronomy at High Energies \\ An Experimental Review}

\Author{Christian Spiering 
 }

\AuthorNames{Christian Spiering}

\address[1]{Deutsches Elektronen-Synchrotron DESY,  
Platanenallee 6, 15738 Zeuthen, Germany; christian.spiering@desy.de}

\abstract{Neutrino astronomy at high energies is an emerging field but still in a state of infancy. It is barely a dozen years ago that a diffuse flux of extraterrestrial neutrinos with energies in the TeV and PeV ranges was detected and even fewer that first individual sources could be identified. A new window to the universe has been opened after four decades of efforts to realize the gigantic instruments that made that breakthrough possible. This review describes the road towards the present, reviews the actual status of  the field and sketches future developments.}

\keyword{neutrinos; neutrino telescopes; cosmic rays; active galaxies; particle acceleration}

\begin{document}

\section{Introduction}
\label{Intro}

With the detection of solar neutrinos in 1968 \cite{Davis:1968cp}
and neutrinos from a supernova in 1987 
\cite{Kamiokande-II:1987idp,Bionta:1987qt,Alekseev:1988gp}, astrophysics has moved beyond the boundary set by electromagnetic radiation. The year 2013 marked the discovery of a diffuse cosmic flux of high-energy neutrinos \cite{IceCube:2013low}, expanding the neutrino window towards TeV energies. Eventually, in 2015, the discovery of gravitational waves \cite{LIGOScientific:2016vlm} added another astrophysical window. Meanwhile, the combination of these different information carriers led to what is called “multimessenger” astronomy, dubbed in analogy to the well-established multi-wavelength astronomy, which combines observations from various electromagnetic wavelength bands.


The first ideas to use neutrinos as cosmic messengers from objects beyond the solar system were expressed in 1960 \cite{Greisen:1960wc,Reines:1960we}. The authors argued that a supernova stellar collapse in our galaxy would be accompanied by a short burst of neutrinos in the 5--20 MeV range. Moreover it was argued that pulsars must accelerate charged particles in their magnetic fields. These particles would hit matter -- either in the source or on their way to Earth -- and generate pions and subsequently neutrinos as their decay products. Meanwhile, other mechanisms are considered to be the main source of high-energy cosmic particles: the acceleration across shock waves -- be it the shock waves of supernova remnants or the shock waves moving along the jets of active galaxies -- to mention just two of several possible~origins.


The instruments to detect these neutrinos (as any natural neutrinos) have to be arranged deep underground or underwater to be shielded against all sorts of background, notably charged particles from cosmic-ray air showers. While five decades ago some hoped that the size of underground detectors might be sufficient to catch a few high-energy neutrinos, it was clear that larger detectors would be necessary to truly launch neutrino astronomy beyond the MeV range typical for solar and supernova neutrinos. This was the motivation to build neutrino detectors in open water, starting with the cubic-kilometer DUMAND project off the coast of Hawaii \cite{Roberts:1992re}. The goal of a cubic-kilometer scale was reinforced in \cite{Barwick:1991ur,Halzen:1993qv} with the claim that {\it ``at the 1 km$^2$ size it seems inescapable, based upon present 
calculations and simple energetic considerations extrapolating from observations with gamma 
rays and lower-energy photons, that such [i.e., neutrino, C.S.] point sources must be seen.''.} 

Section \ref{Why} of this review summarizes the arguments for conducting neutrino astronomy. Section \ref{NTs} describes the operating principles of large neutrino detectors, focusing on detectors that record the Cherenkov radiation emitted by secondary particles generated in neutrino interactions. It also sketches the emergence of these detectors (for a detailed historical review, see \cite{Spiering:2012xe}). The focus of this review is on the astrophysical results (status May 2026), which are discussed in Sections \ref{diffuse-section}--\ref{point-source-section}. Section \ref{diffuse-section} presents the results on the diffuse flux of high-energy cosmic neutrinos, and Section \ref{GP-section} presents the results on neutrinos from the Galactic Plane. The results on point sources are reviewed in Section \ref{point-source-section}. A timeline with the most important astrophysics results described in Sections \ref{diffuse-section}--\ref{point-source-section} is given in Table \ref{tab1}. 

\begin{table}[H]
\caption{
Timeline of 
 the main results obtained with large neutrino telescopes (status May 2026). The term “discovery” is usually used if the significance level is 5 standard deviations or higher. The first observations of a diffuse all-sky flux and galactic neutrinos were a bit below the 5\,$\sigma$ 
 value, but the scale of the observed signal was in agreement with predictions. Meanwhile the corresponding significance values are more than 7\,$\sigma$ and 5\,$\sigma$, respectively. For signals larger than 4 but clearly below 5$\sigma$, the term “evidence” is used.}\label{tab1}
 \begin{tabularx}  {\textwidth}{Llll}
\toprule
\textbf{Result}	& \textbf{Detector} &	\textbf{Year} &	\textbf{Section} \\
\midrule 
Discovery of a diffuse flux of cosmic neutrinos	& IceCube &	2013 & Section 
 \ref{sec4.1} \\
\midrule
Evidence for a transient point source, & & & \\
the blazar TXS 0506-056	& IceCube	& 2017 &	Section \ref{transient-subsection} \\
\midrule
Detection of a shower at the Glashow resonance &	IceCube	& 2021	& Section \ref{sec4.3}\\
\midrule
Evidence for a steady point source, & & & \\
the Seyfert galaxy NGC 1068	& IceCube & 2022 &	 Section \ref{sec6.1} \\
\midrule
Discovery of neutrinos from the Galactic Plane &	IceCube &	2023 &	Section \ref{GP-section} \\
\midrule
Confirmation of the diffuse flux of cosmic neutrinos &	Baikal-GVD &	2025 &	 Section \ref{sec4.1} \\
\midrule
Detection of a neutrino event with & & &  \\
a most likely energy of 220 PeV	& KM3NeT &	2025 &	 Section \ref{sec4.3} \\
\midrule
Present best upper limit in the \linebreak 10 PeV--10 EeV region	& IceCube	& 2025 &	Section \ref{sec4.4} \\
\bottomrule
\end{tabularx}
\end{table}

\vspace{-2pt}
Section \ref{Synergies-section} sketches some synergetic effects of cooperation among neutrino telescopes on the one hand and
between neutrino telescopes and telescopes for electromagnetic radiation on the other hand. 
Section \ref{Conclusions-section} gives a short summary.

I will not discuss particle physics aspects in this review, although they account for a considerable part of the scientific program of neutrino telescopes. 
For a review of neutrino interaction physics in neutrino telescopes I refer the reader to  \cite{Katori:2021nwq}; for searches for beyond-standard-model physics with astroparticle physics instruments to \cite{Ackermann:2023gmd}; and for dark-matter searches with neutrino telescopes to \cite{deDiosZornoza:2021rgw,Gozzini:2025ivt}.

Also, phenomenological modeling of sources of high-energy neutrinos will not be discussed in any detail. 
For reviews with a stronger emphasis on theoretical and phenomenological aspects, I refer the reader to
\cite{Berezinsky:2010fqt,Fiorillo:2024jqz,Arguelles:2024ncf,Troitsky:2021nvu,Winter:2024zpk,Groth:2025aan}.

\section{Why Neutrino Astronomy?}
\label{Why}

The basic rationale for neutrino astronomy is finding and understanding the sources of high-energy cosmic rays, the spectrum of which extends beyond $10^{20}$ eV \footnote{Actually, cosmic rays have been the first cosmic messenger beyond visible light. They were discovered in 1912 \cite{Hess:1912srp} and have been studied since then with detectors of gradually increasing complexity and size.}.  
Charged particles, however, are deflected by large-scale cosmic magnetic fields and do not point back to their sources. The deflection is inversely proportional to the energy and charge of the particle. For the highest energies, above a few $10^{19}$ eV, protons propagating in our galaxy are deflected by only a few degrees. But even the world's largest cosmic ray detector, the Pierre Auger Observatory, has not yet succeeded in clearly locating individual sources. This provides a clear argument to search for gamma rays and neutrinos from these objects. They are generated in interactions of cosmic rays with ambient matter,
\begin{equation}
    p(A)+p(A,\gamma)\to\pi +X
\label{CR-interactions}
\end{equation}

\noindent
(with $A$ being a nucleus and $X$ a hadronic system), followed by the decays

\vspace{2mm}

\begin{equation}
 \pi^+\to\mu^+ + \nu_\mu   
    \quad\text{and}\quad 
    \mu^+\to e^+ + \bar{\nu}_\mu + \nu_e\
    \label{piplus-decay}
\end{equation}
\begin{equation}
 \pi^-\to\mu^- + \bar{\nu}_\mu   
    \quad\text{and}\quad \mu^-\to e^- + \nu_\mu + \bar{\nu}_e\
    \label{piminusdecay}
\end{equation}
\begin{equation}
 \pi^0\to\gamma + \gamma    
  \label{pizero-decay}
\end{equation}

\vspace{2mm}

For photo-production, i.e., interactions of protons on ambient photon fields where kinematics prefers final states with only one pion, equation 1 simplifies to

\vspace{2mm}

\begin{equation}
p + \gamma\to\pi^0 + p
    \quad\text{\   and\   }\quad
    p + \gamma\to\pi^+ + n
\label{photon fields}
\end{equation}

\vspace{2mm}

Here, neutrinos and gamma rays are produced at roughly equal rates.
About $20\%$ of the proton energy is transferred to the pion. 
A neutrino
carries about $5\%$ of the proton~energy.

The co-production of gamma rays and neutrinos raises the question of why not only the much easier-to-detect gamma rays are used to learn more about the sources of cosmic rays. Actually, there are several arguments in favor of neutrinos.

Firstly, gamma rays could emerge in purely electromagnetic processes where electrons have been accelerated, not protons (which make up $99\%$ of charged cosmic rays). The electrons can produce high-energy gamma rays via Compton scattering, e.g., with ultraviolet or X-ray photons.

\vspace{2mm}

\begin{equation}
e^- + \gamma_{\text{low energy}}\to e^- + \gamma_{\text{high energy}}
\label{Compton}
\end{equation}

\vspace{2mm}

Therefore, incontrovertible evidence for hadron acceleration can be obtained only from neutrinos rather than from gamma rays. Simultaneous data on the ambient photon fields and the high-energy gamma-ray spectra can provide {\it indirect} evidence for one or the other hypothesis, but certainly not something that definitely proves the absence or presence of hadrons accelerated in the source \footnote{In fact, almost three hundred gamma-ray sources with TeV energies, some even with PeV energies, have been detected. A close look at the gamma-ray spectra suggests that, for most of these sources, the proportion of gamma quanta from $\pi^0$ decays plays no or only a minor role.}.

Secondly, many sources of high-energy gamma rays are likely located in an extensive cocoon of matter or photon fields \cite{Berezinsky:1980mh}. Actually, powerful neutrino sources require an efficient (and that means dense) target to produce pions. This cocoon is likely to be opaque to the gamma rays from $\pi^0$ decay that accompany the neutrinos. On their way through this cocoon, the gamma rays are scattered many times and release energy during each scattering process so that they reach the outside only with MeV or GeV energies. If the cocoon is thick enough, only photons in the infrared range escape the cocoon. 

Thirdly, gamma rays interact with the extragalactic background light (EBL), which is due to star formation processes, and at the highest energies with 2.7 K cosmic microwave background (CMB). The process $\gamma + \gamma_{\text{EBL/CMB}}\to e^+ + e^-$ initiates an electromagnetic cascade, and the initial energy of the gamma ray is showered down to MeV or GeV energies.

\begin{figure}
{
\includegraphics[width=14 cm]{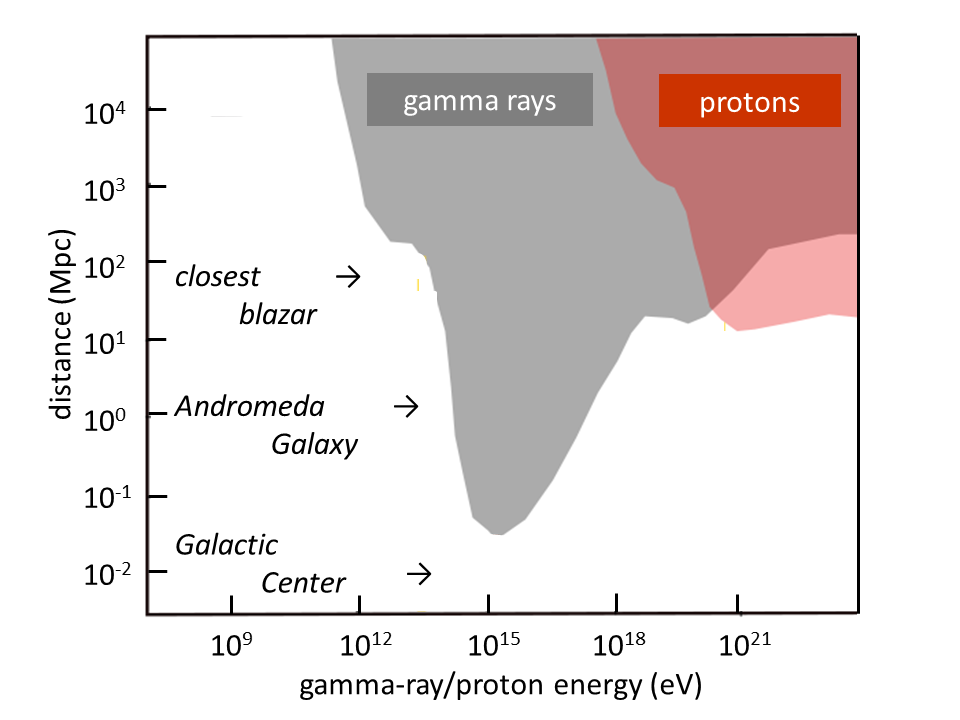}
\caption{Range of 
 gamma rays and protons due to interactions with EBL and CMB radiation, respectively. The colored regions mark the obscured regions (gray for gamma rays; red for protons). The range of protons is limited by the GZK effect ($p +\gamma_{\text{CMB}} \to \Delta$). Modified after \cite{Batten:2022ajr}. \label{range}}
}
\end{figure}

Figure~\ref{range} sketches the range of gamma rays and protons in cosmic space. The steep drop at $\sim5\,\times\,10^{13}$ eV is due to interactions with CMB photons. For target photons with higher energy, like EBL photons, the process already sets in at a few TeV. Due to the lower cosmic density of EBL photons compared to CMB photons, the effect appears only for longer distances than for the CMB case. 

As can be seen from the figure, above a few tens of TeV, the sole long-range neutral information carriers are neutrinos, while, above one TeV, only 1\% of the visible universe and, at $\sim\,1$ PeV, only our own galaxy can be explored with gamma rays. 
Note that, due to interactions with the CMB, the range of protons is also limited.

\vspace{2mm}
\section{Neutrino Telescopes}
\label{NTs}
\vspace{1mm}
\subsection{Optical Detection: Basics}
\label{Optical-NT}

This subsection focuses on neutrino telescopes, which detect the Cherenkov light from secondary particles generated in neutrino interactions (see also U. Katz and C.S. in~\cite{ParticleDataGroup:2024cfk}). These telescopes are three-dimensional arrays of ``optical modules'' (OMs) -- glass spheres installed in open transparent media like water or ice. They are arranged at depths that completely block the
daylight as well as most of the muons emerging from cosmic air showers in the atmosphere above the telescope. The OMs contain photomultipliers that are sensitive to single photons. They are arranged along vertical strings, with distances of 15--30~m between the OMs at one string and horizontal distances of 50--150 m between the strings. The typical energy range targeted by these instruments is $E_\nu\gtrsim100\,$GeV. Dedicated setups with smaller distances can address lower energies down to a few GeV (see Section~\ref{opticalNTs}).

The arrival time of photons at the OMs is registered with a precision of a few nanoseconds. The time pattern allows the reconstruction of the direction with sub-degree accuracy. The hit pattern also enables reconstructing the neutrino energy.




\subsubsection{Event Patterns}

\begin{figure}[H]
{
\includegraphics[width=14.0 cm]{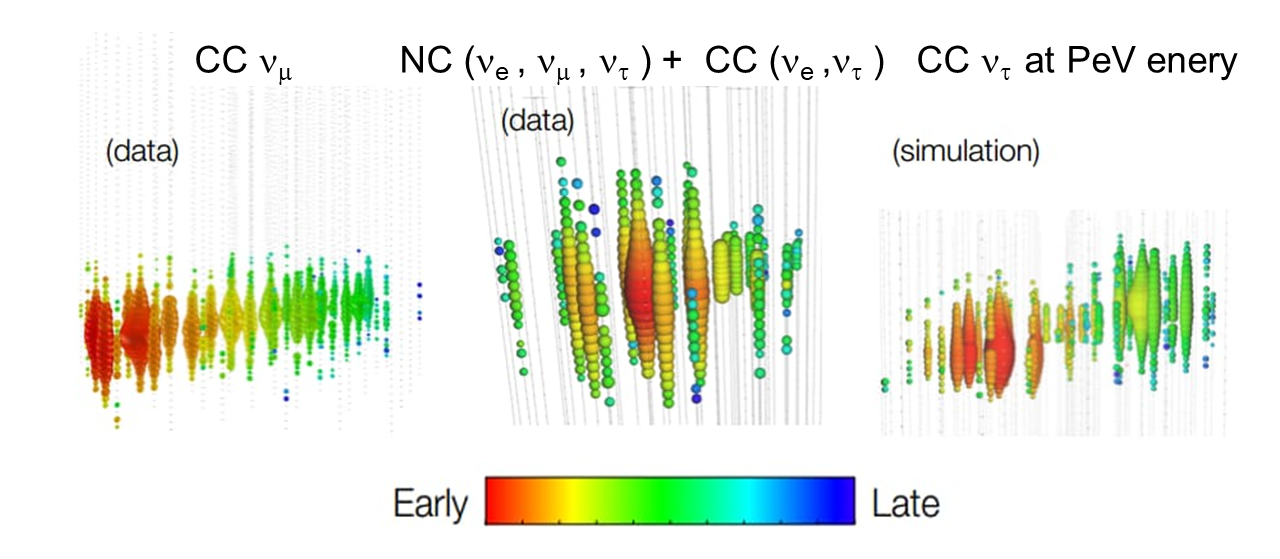}
\caption{Example for the three event signatures in the IceCube Neutrino Telescope (see below): an upgoing muon track from a $\nu_{\mu}$ charged current interaction outside the detector (\textbf{left}), a cascade event from an interaction within the detector (\textbf{center}) and a Monte Carlo-simulated double-bang event from a 
$\nu_{\tau}$ interaction within the detector (\textbf{right}). Each hit optical module (OM) is marked by a bubble. The larger the diameter, the larger the light signal. The color code marks the light arrival time at the OM. \label{signatures}}
}
\end{figure}

A charged current (CC) interaction of a muon neutrino produces
a muon track and a hadronic particle cascade. All the neutral current (NC) reactions as well as CC reactions of electron neutrinos produce particle cascades only. CC interactions of tau neutrinos can lead to either signature depending on the $\tau$ decay mode. A $\tau$ decay sufficiently far away from the primary interaction (at PeV energies and above) can produce a second cascade that is distinguishable from the primary vertex. The result is a third pattern called double-bang signature. Figure~\ref{signatures} sketches the three cases. 

\vspace{2mm}
\subsubsection{Angular Resolution and Medium Properties}

The average mismatch angle between the final-state lepton and the neutrino is about
$\langle{\phi_{\nu\ell}}\rangle\sim1^\circ/(E_\nu/{\rm TeV})^{0.5}$, with a slightly
steeper decrease beyond 10\,TeV. 
This defines the {\it intrinsic} 
 kinematic limit to the angular resolution. The reconstruction accuracy for the final state lepton can be much larger. For instance, for CC muon neutrino reactions (Figure~\ref{signatures}~left),  the angular resolution above 10\,TeV is dominated
by the muon reconstruction accuracy. It improves with the amount of light collected (which increases with energy above a TeV). 
Light propagation in the medium has a significant impact on the angular accuracy: absorption reduces the light signal and
scattering smears the arrival time, both worsening the angular resolution.

The values for the peak absorption lengths in sea water, best lake waters (Lake Baikal) and deep Antarctic ice (IceCube) are about 50\,m, 20\,m and 100\,m, and those for the ``effective scattering length'' (the scattering length divided by ($1-\cos{\theta}$), with $\theta$ being the average scattering angle), are 200\,m, 200\,m and 20--40\,m, respectively \cite{Spiering:2020iih}. 

\begin{figure}
{
\includegraphics[scale=0.4]{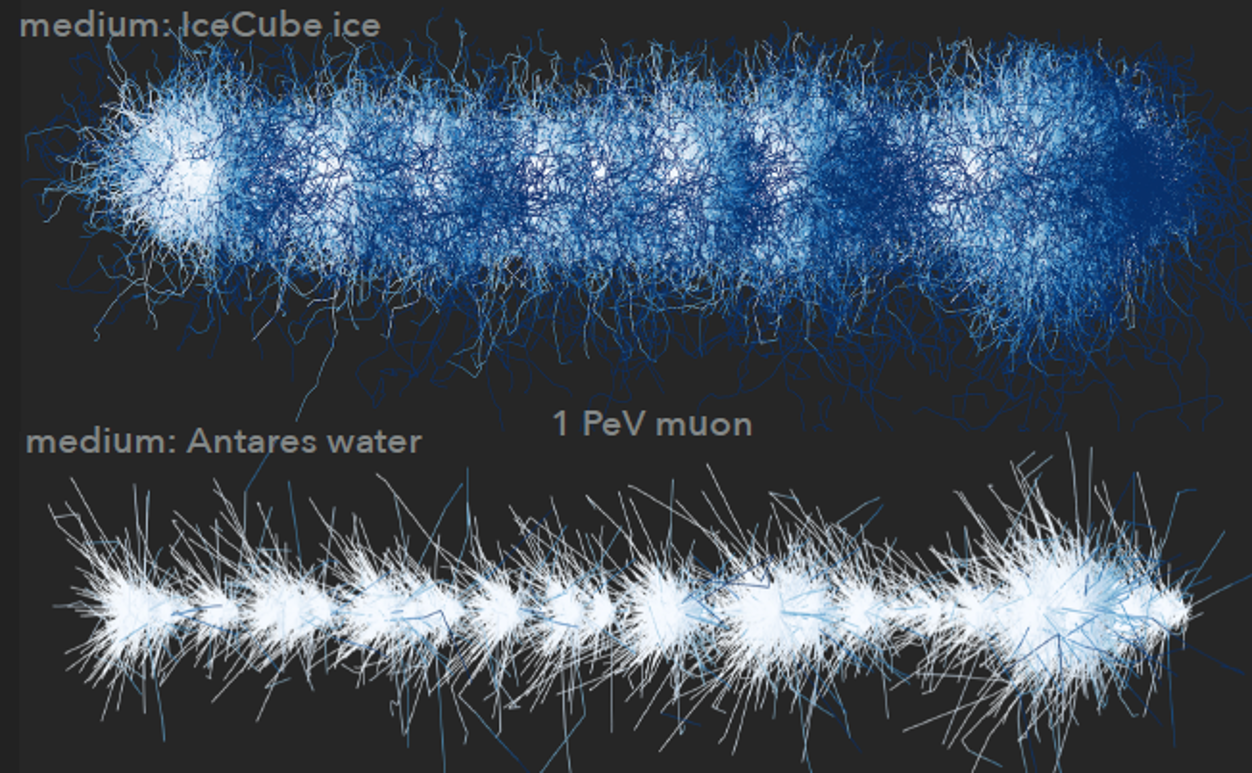}
\caption{Light emission along a 1-PeV muon moving from right to left. \textbf{Top}: Antarctic ice (IceCube); \textbf{bottom}: Mediterranean water (ANTARES site). At these energies, light emission is dominated by emission from cascades (pair production, bremsstrahlung, and photo-nuclear interactions) along the track. The color encodes the time after photon emission, white for ``young'' photons,  blue 
 for ``old'' photons. While photons in water propagate rather straight, photons in ice are strongly scattered --- a drawback for precise timing. On the other hand photons in ice live longer --- an advantage for the amount of detected light. Figure courtesy of Kai Krings, TU Munich.}
\label{Light-emission}}
\end{figure}

Figure \ref{Light-emission} illustrates the different light propagation in water and ice.
It demonstrates that polar ice is best in light collection (many ``old'' photons in blue) and worst in the ``quality'' of  light regarding pattern recognition. Lake water, due to the small absorption length, is worst in light collection and requires a denser instrumentation than sea water to collect the same amount of light. Sea water is a bit worse than ice in terms of light collection due to the shorter lifetime of photons in water.

\begin{figure}
{
\includegraphics[scale=0.70]{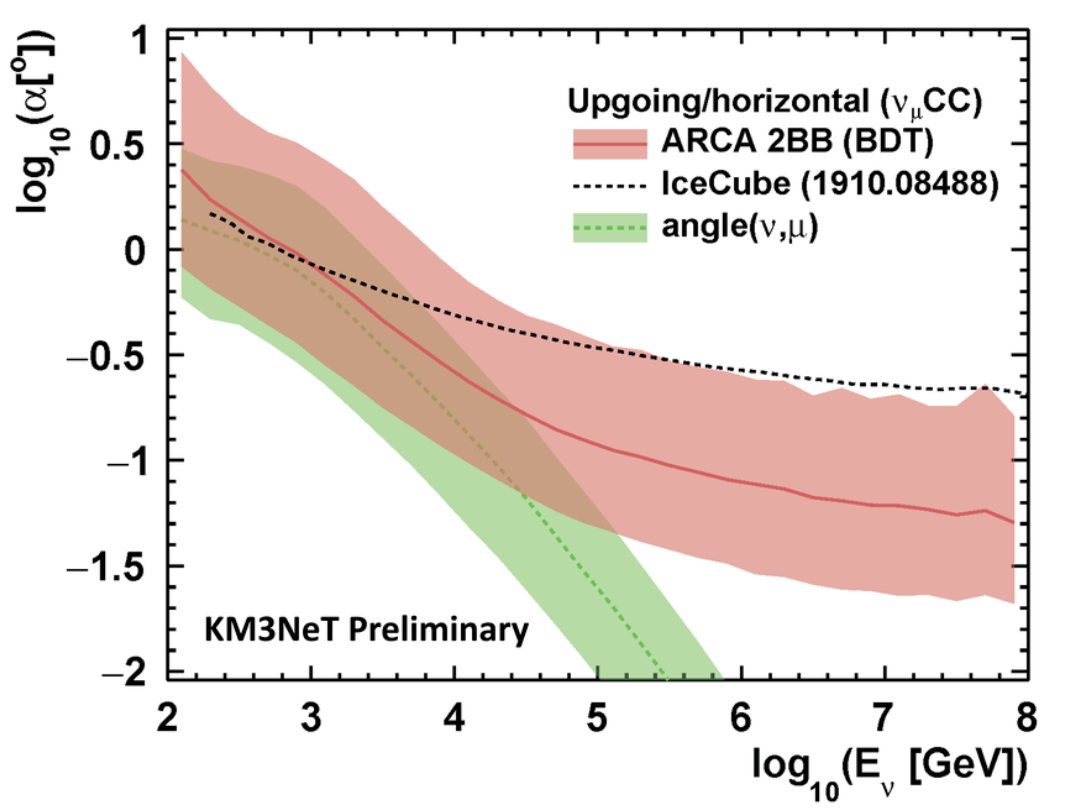}
\caption{Expected angular resolution as a function of neutrino energy for $\nu_{\mu}$ charged current events in KM3NeT (IceCube curve from \cite{IceCube:2019cia}). Figure taken from \cite{KM3NeT:2021szv}.}
\label{angres}
}
\end{figure}

Figure \ref{angres} 
compares the MC-calculated angular resolution for muon tracks in KM3NeT and IceCube. At 1 PeV muon energy, the resolution reaches $\sim$0.1 (0.3) degrees for KM3NeT (IceCube), the inferior IceCube value being due to the stronger light scattering in ice.

\unskip

\subsubsection{Energy Determination}

For muon energies $E_\mu\gtrsim1\,$TeV, the light emission increases due to
radiative processes like pair production or bremsstrahlung. This enables estimating $E_\mu$ from the recorded light intensity with an accuracy of $\sigma(\log{E_{\mu}})\sim$ 0.3--0.5.

\begin{figure}
{
\includegraphics[width=0.9\textwidth]{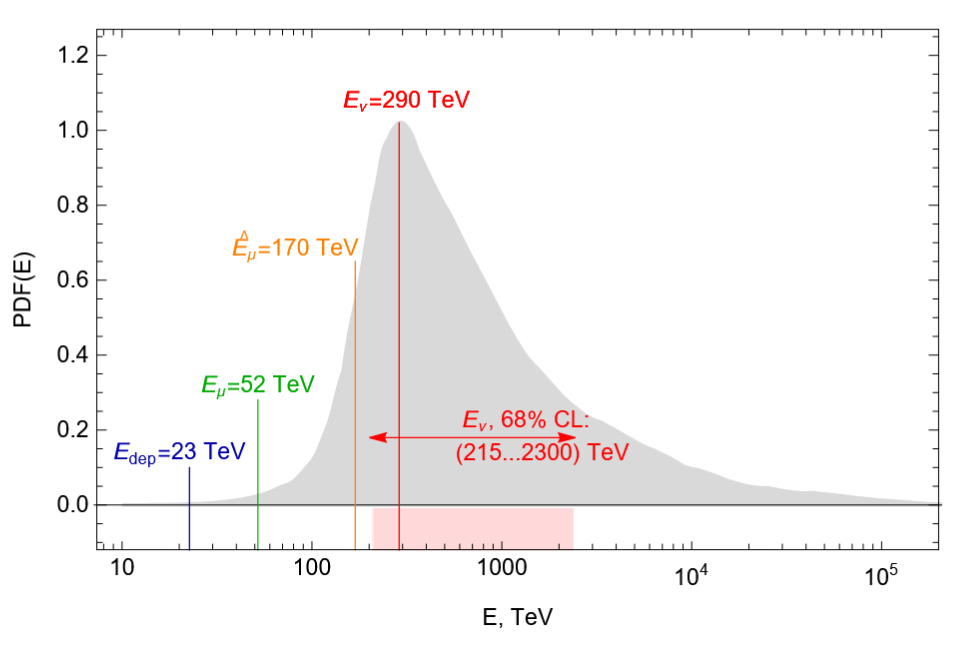}
\caption{
An illustration 
 of the energy-estimate uncertainty
for a track event using the example of one of the best known
neutrinos registered by IceCube (IC170922A, coincident with
the outburst of the blazar TXS 0506+056; see Section \ref{point-source-section}).
The horizontal axis displays energies: the energy deposited in the detector 
$E_{dep} = 23.7 \pm 2.8$ TeV, the reconstructed energy of the
muon entering the detector \linebreak $E_{\mu}$ = 52 (+11, $-$9) TeV, the energy estimate 
of the muon at birth $\hat{E}_{\mu}$ = 170 TeV, and the most likely
neutrino energy $E_{\nu} \simeq$  290 TeV. The shaded area shows the
probability density function (PDF) of the values of $E_{\nu}$; also
shown in red is the uncertainty region of the  $E_{\nu}$  values --- from 215 to
2300 TeV (68\% CL). A neutrino power-law spectrum 
with a spectral index of 2.13 is assumed. Plotted on the
basis of data from \cite{IceCube:2018dnn}. Figure taken from
\cite{Troitsky:2021nvu}.
}
\label{TXS-energy-uncertainties}}
\end{figure}

The drawback of throughgoing muons is that one does not know at which distance to the telescope they have been generated, i.e., how much energy the muon has lost already. The energy of the muon at the moment of creation and the neutrino energy can only be estimated statistically. While the neutrino energy is constrained
from below quite well (it cannot be lower than the energy of the muon entering the detector), the broad non-Gaussian distribution of possible neutrino energies extends towards values orders of magnitude
higher; see Figure~\ref{TXS-energy-uncertainties} taken from \cite{Troitsky:2021nvu}. Note that the probability distribution and the most likely value of the neutrino energy depend on the assumption about the spectrum of
astrophysical neutrinos.

If the muon is starting and stopping in the detector (i.e., for rather low muon energies), $E_\mu$ can be estimated from the length of the track. 

\subsubsection{Tracks vs. Cascades}

The better angular resolutions compared to cascade events make CC $\nu_\mu$
reactions the preferred channel to identify point sources.
If one is interested in the most precise energy measurement, the cascade channel (Figure~\ref{signatures}, center) is the best option. Particle cascades at energies discussed here are 5--20\,m long, i.e., short compared to OM distances. For cascades that are fully contained in the instrumented volume, their energy can be determined from the total amount of recorded Cherenkov light ---  with
about $20\%$ accuracy at energies above 10\,TeV and $10\%$ beyond 100\,TeV.
The directional accuracy for cascade events, however, is significantly worse than for muon tracks. Above 100\,TeV, the average angular error of cascades is about $2^{\circ}$ in water and $10^{\circ}$ in polar ice, the difference being due to stronger light scattering and the inhomogeneity of 
ice. Recent improvements in the reconstruction algorithm, in particular the inclusion of birefringence effects, have improved the angular accuracy of cascades in IceCube to about $5^{\circ}$ \cite{IceCube:2024csv}.

Cascade events have a clear advantage in identifying a high-energy excess of extraterrestrial over atmospheric neutrinos in the diffuse flux. Despite their poorer angular resolution, however, they can also complement track
events in searches for {\it transient} signals from point sources where the background from atmospheric neutrinos is small and the angular search window can be kept larger than for steady sources.

\subsubsection{Transparency of Earth}

Regarding the throughgoing muon signature, neutrino telescopes are sensitive to the opposite hemisphere (including, depending on how deep the detector is, $\sim$$10^\circ$ above the horizon). At energies above a few TeV, however, neutrino absorption in Earth becomes noticeable. For vertically upward-moving neutrinos (zenith angle $\theta=180^\circ$), the survival probability is 74\%\hl{} 
 (27\%, <2\%) for 10 (100/1000)\,TeV. The search window for PeV neutrinos, therefore, is narrowed to a small region close to the horizon. Looking further upward, the background from atmospheric muons becomes dominant. 

Sensitivity to neutrinos of all flavors and from all directions can be
achieved by selecting events that start inside the
instrumented volume and have no early hits in
the outer layers of the detector (i.e., no tracks entering the detector from outside).
Naturally, such a veto rejects downgoing  muons, 
but it also rejects downgoing {\it atmospheric neutrinos} that are
accompanied by muons from the same air shower and thus reduces the
atmospheric-neutrino background \cite{PhysRevD.79.043009}.

\subsubsection{Effective Area}

\begin{figure}[H]

\includegraphics[scale=0.75]{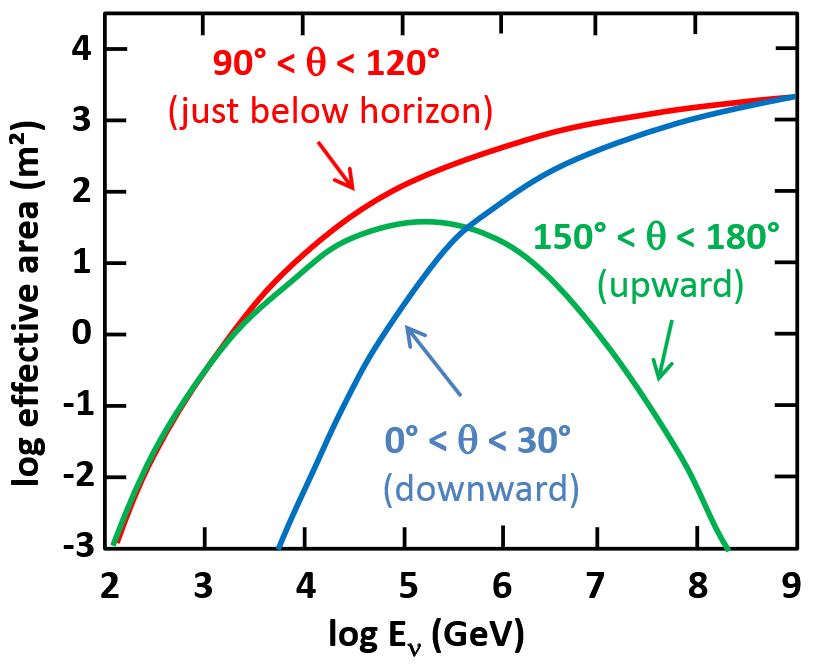}
\caption{Average 
 over the effective areas for IceCube as an example of a cubic$-$kilometer neutrino telescope as a function of neutrino energy for three intervals of the zenith angle $\theta$.  The reduction with energy at large zenith angles (green curve) is due to Earth absorption; the onset with energy at angles above the horizon (blue curve) is due to the decreasing background from atmospheric muons with increasing energy.  The values shown here correspond to a specific event selection for point source~searches.}
\label{fig6area}
\end{figure}

The sensitivity of a neutrino telescope is quantified by its ``effective area'' --- the
fictitious area for which the full flux of incoming neutrinos would be recorded.
The effective area rises with the neutrino energy due to the rise of neutrino cross section and muon range.
Given the minuscule neutrino cross section, the effective area is many
orders of magnitude smaller than the geometrical area of the detector. For a cubic-kilometer neutrino telescope it rises from 10 cm$^2$ at 100 GeV to 1 m$^2$ at some TeV and
100 m$^2$ at 100 TeV; see Figure~\ref{fig6area}. 
Translated to a detection efficiency, this means that a 1\,TeV muon neutrino would be detected with a probability of the order $10^{-6}$ if the telescope is on its path.

\subsection{Past, Present and Future Neutrino Telescopes with Optical Detection}
\label{opticalNTs}

Ideas to detect high-energy cosmic neutrinos with detectors underground or underwater date
back to the late fifties. In 1960, Kenneth Greisen and Frederick Reines discussed motivation
and prospects for such devices \cite{Greisen:1960wc,Reines:1960we}. In the same year,
Moisei Markov proposed  {\it ``...to install detectors deep in a lake or a sea
and to determine the direction of charged particles with the help of Cherenkov radiation”} 
\cite{Markov:1960vja}. Actually, this approach
appeared to be the only way to build detectors beyond the scale of $10^4$ tons. 

The march towards underwater neutrino telescopes started in 1973 with the idea
to build a {\it Deep Underwater Muon and Neutrino Detector} (DUMAND) \cite{Roberts:1992re}. 
Five years later, an array of about 20\,000 photomultipliers spread over a 1.26 cubic-kilometer volume was proposed. 
Technological and financial constraints forced the protagonists to reduce the size of this gigantic device in several steps, down to a 216-PMT version (DUMAND-II, 1988). The detector was to be installed close to Hawaii at a depth of 4.8\,km. After several promising deployments of test devices, however, the installation of a first string failed. Despite the remarkable progress over the years, the project was terminated in 1995.

In 1980, Alexander Chudakov proposed to use Lake Baikal in Siberia as the site
for a ”Soviet DUMAND”. In late winter the lake is covered by a thick ice layer, which
can be used to deploy underwater devices without the need for ships. 
Soon, a site in the southern part of the lake, with a depth of about 1370\,m at 3.6\,km from shore, was found
and prototypes were tested.  After a decade of development, a detector named NT200 was completed in 1998. It comprised 192 OMs at 8 strings \cite{BAIKAL:1997iok}. The modules housed PMTs with 37\,cm diameter. The first two neutrino candidates (i.e., upward moving muons) were identified in the 1994 data, taken with a partial configuration of three strings, each carrying 12 OMs. Together with AMANDA (see next paragraph) NT200 provided the first proof of principle for underwater neutrino telescopes despite its geometrical volume of only $\sim$$10^{-4}$\,km$^3$. 

The second DUMAND-type detector was AMANDA ({\it Antarctic Muon and Neutrino Detector Array}), based on an idea suggested by Francis Halzen and John Learned \cite{Andres:1999hm}. AMANDA was installed in the 3\,km thick ice shield covering the South Pole. The holes to contain OMs were melted with hot water, which froze back within some days after the strings with OMs had been lowered. OMs were arranged  at 1500--2000 m depth. AMANDA comprised 677 OMs at 19 strings, each containing an 8-inch PMT pointing downwards.  The geometrical volume of AMANDA was 0.015 km$^3$. Construction started in 1995, and the detector was completed in January 2000. It recorded almost 7000 neutrino candidates over eight years. 

The first neutrino telescope in the sea was ANTARES ({\it Astronomy with a Neutrino Telescope and Abyss environmental RESearch}) \cite{ANTARES:2011hfw}. It was constructed close to Toulon in the 
Mediterranean Sea at a depth of nearly 2500\,m and 
completed in 2008. ANTARES consisted of 12 strings, instrumented over 350\,m length with 25 “stories”. A 
story was equipped with three 10-inch PMTs housed in 13-inch glass spheres. The PMTs were oriented at 45° 
with respect to the vertical. The detector, with its geometrical volume of 0.010 km$^3$, was operated until 2021 and registered about 9000 neutrino candidates.

The characteristics of the neutrino events recorded by all three first-generation detectors were consistent with an atmospheric origin. Only the ANTARES data have shown a slight indication of an excess at the highest energies that could possibly be assigned to extraterrestrial neutrinos 
\cite{ANTARES:2024ihw} --- a decade after IceCube had detected a diffuse flux of cosmic neutrinos (see below). 
Not the slightest hint was found for a point source of neutrinos. It was clear that much larger detectors would be necessary to identify the contribution of extraterrestrial neutrinos. This made the case for three neutrino telescopes on the cubic-kilometer scale: IceCube \cite{IceCube:2016zyt}, KM3NeT \cite{KM3Net:2016zxf} and Baikal-GVD \cite{Baikal-GVD:2025iwm}.

\begin{figure} [H]
\includegraphics[width=.99\textwidth]{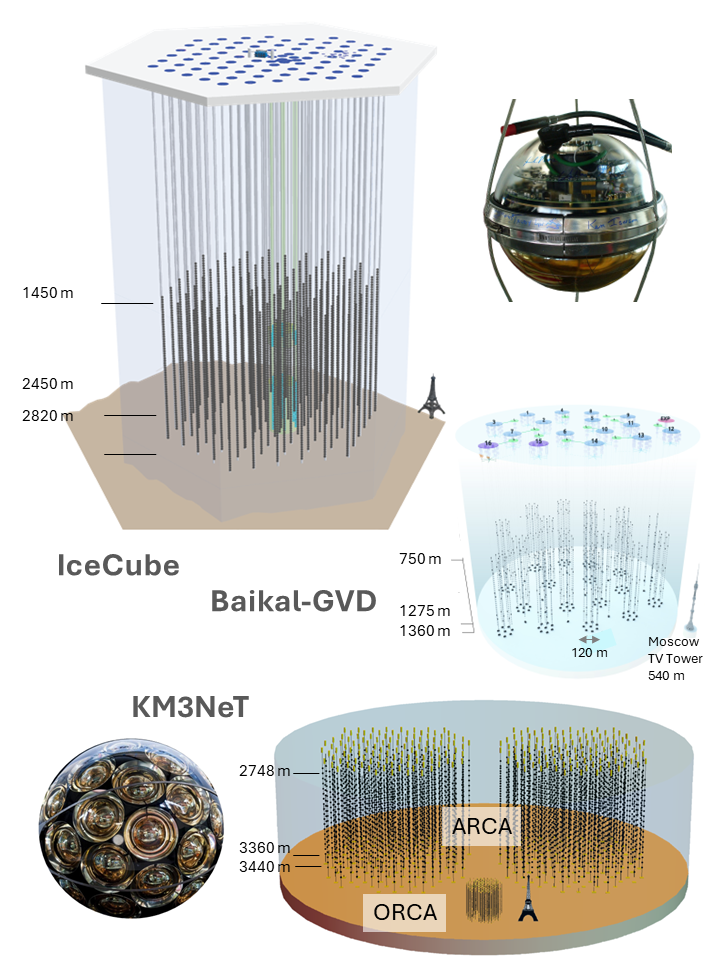}
\caption{The three 
 presently operating neutrino telescopes, IceCube, Baikal-GVD and KM3NeT. IceCube is shown in its 2010--2025 configuration, Baikal-GVD in its May 2026 configuration with 16~clusters and one ``experimental cluster'' to test new technologies (top right in the figure), and KM3NeT in its full configuration to be reached in a few years (as of May 2026, 51 of the \mbox{$2\times115$}~strings of KM3NeT and 38 of the 115 strings of ORCA are installed).
The shown relative sizes of the detectors are only approximate. The depth information left of the detectors gives the depths of the top and bottom optical modules and that of the bed rock/sea ground of IceCube, 
Baikal-GVD and KM3NeT ARCA. The corresponding values for KM3NeT-ORCA are 2247\,m, 2410\,m and 2450\,m. }
\label{detectors}
\end{figure}

IceCube (see Figure~\ref{detectors}, top) (built 2004--2010) consists of 5160 digital OMs
(DOMs) installed on 86 strings at depths of 1450 to 2450\,m in Antarctic
ice. String distances are 125\,m and the vertical
spacing between OMs 17\,m. A high-density sub-array {\it DeepCore} is
arranged at the center of IceCube at depths where the ice properties are best. 
A further 324 DOMs are installed in IceTop, an array
of detector stations on the ice surface above the strings. Each DOM contains a downward-pointing 10-inch PMT.  
In January 2026, {\it IceCube-Upgrade} was installed.  
It consists of five densely equipped strings and further densifies the DeepCore region. While the threshold of the full IceCube array is about 100\,GeV, that of DeepCore is about 10 GeV and that of the upgrade about 2 GeV. While IceCube/DeepCore DOMs contain single 10-inch PMTs (see Figure~\ref{detectors}, top right), part of the upgrade DOMs contain 25~PMTs with 7.5\,cm diameter (following the KM3NeT mDOM idea; see below) and  another part two 20\,cm PMTs pointing back to back. 

KM3NeT will consist of blocks of 115 strings each, with 18 DOMs per
string. The collaboration envisages two 
blocks for neutrino astronomy, with vertical spacings between OMs of 36\,m and
string distances of 90\,m (ARCA, for {\it
Astroparticle Research with Cosmics in the Abyss}), and one single block to
determine the neutrino mass hierarchy, with OM spacings
of 9\,m and string distances of about 20\,m (ORCA,
for {\it Oscillation Research with Cosmics in the
Abyss}); see Figure~\ref{detectors}, bottom right. ARCA is being installed near Capo
Passero, east of Sicily (depth 3440\,m), and ORCA near Toulon (depth 2450\,m). 
In May 2026, 55/38~strings have been
deployed for ARCA/ORCA, and 49/38 of them are continuously taking data.  Completion of the full
ARCA and ORCA arrays is planned for 2030.

A novel concept has been chosen for the OMs of
KM3NeT (see Figure~\ref{detectors}, bottom left). Each OM (diameter 43\,cm) houses 31 PMTs (7.5\,cm). This \textit{mDOM} (multi-PMT DOM) has several advantages: (1) the total
photocathode area exceeds that of a single 25 cm PMT by more than a factor of ~3; (2)
the individual readout of the PMTs results in a better separation between
one- and two-photoelectron signals, which is helpful for online data filtering
and random background suppression; (3) the hit pattern of a single OM provides
directional information; (4) no mu-metal shielding against Earth magnetic
field is required \footnote{For large hemispherical PMTs, effective and uniform focusing of photoelectrons to the first dynode is a problem due to Earth's magnetic field. It can be overcome by a mu-metal wire grid covering the PMT.}.

In Baikal-GVD, the strings are arranged in clusters (see Figure~\ref{detectors}, center). Each cluster comprises 8 strings --- a central
one and 7 peripheral strings --- across a radius of ~60 m.  Each string carries 36 DOMs spaced 15\,m apart and containing a single 25 cm PMT, similar to IceCube.
Several clusters also have an additional ``inter-cluster'' string that is placed in the otherwise non-instrumented space
between clusters. The distance between the cluster centers varies between 250 and 300 m.  As of May 2026,
16 full clusters (4860 DOMs) have been installed and are currently taking data. 
While over the next few years Baikal-GVD clusters with their present technical design will be deployed, concepts for a much larger array are under discussion.

The photomultiplier data are digitized underwater/ice: for IceCube and KM3NeT inside the OMs, for Baikal-GVD in special modules that receive the analog signals from the OMs via maximally 90 m long coaxial cables. The digitized data from IceCube OMs are transmitted via electrical cables to the surface (3.3 km length for the lowest OMs). KM3NeT and Baikal-GVD use optical fibers to transmit the digitized data to shore (several 10 km for KM3NeT and about 4 km for GVD).

For IceCube pulses appearing in local coincidences, the fully digitized waveform is transmitted---for pulses in isolated OMs only time and charge values. For KM3NeT, time and time-over-threshold values for {\it all} PMT signals above an adjustable threshold are sent to shore (i.e., without any local trigger, enabling maximum flexibility at shore). The situation is more complicated and restrictive in the case of GVD: The local coincidence condition requires amplitudes of $>4.5$ and $>1.5$ photoelectrons in two neighbored OMs. These high threshold levels are a consequence of the system architecture with its maximum rate of $\approx$300 events/sec. 

The data sent to shore/surface are processed on computer farms, applying multiplicity- and topology-driven filter algorithms to select event candidates. The filter output data rate for IceCube and for a KM3NeT block is of the order 1 TByte/day. 100 GByte/day of IceCube data are transferred via satellite to the Northern Hemisphere.


Several 
 other detectors are in the R\&D phase or planned and are included in Table~\ref{tab2}. Note that the size of future detectors is subject to funding. Moreover, the instrumented volume is only one of several parameters defining the capability of the telescope. String spacing influences the energy threshold, depth the background rejection capabilities, medium quality both energy threshold and angular resolution, etc.
 \begin{table}[H]
\caption{Present and future NT projects.  The milestones
give the years of first data taking with partial
configurations and detector completion. The
size refers to the instrumented volume. The status of projects under construction is as of May 2026.} \label{tab2}
\footnotesize
\begin{tabularx}{\linewidth}{@{}llLlL@{}}
\toprule
\textbf{Experiment } & \textbf{Milestones}           &\textbf{ Location}        & \textbf{Size  (km\textsuperscript{3})}   & \textbf{Remarks}\\

            \midrule

IceCube     &  2005/2010    & South Pole     & 1.0      &   86 strings \\
\midrule
IceCube-Upgrade      &  2026/2026    & South Pole     &          & 5 dense strings in center of~IceCube\\

\midrule
Baikal-GVD  &  2015/--      & Lake Baikal     & ca.~1    & data taking with 16 of 20 clusters\\                                    
\midrule
KM3NeT/ARCA &  2021/--      & Mediterr.\ Sea near Sicily     & ca.~1    & data taking with 49 of 236 strings  \\

\midrule
KM3NeT/ORCA &  2020/--      & Mediterr.\ Sea   near Toulon     & 0.007    & data taking with 38 of 118 strings \\

\midrule
P-ONE &               & Pacific Ocean  & $\cal{O}$(1) & prototype phase\\
\midrule
TRIDENT &             & South China Sea & $\cal{O}$(10) & R\&D phase\\
\midrule
HUNT &        & South China Sea   or Lake Baikal    &$\cal{O}$(10)& R\&D, prototype tests Lake Baikal \\

\midrule
NEON   &    & South China Sea & $\cal{O}$(10) & R\&D phase\\
\midrule
IceCube-Gen2
            &        & South Pole      &  8    & Planned extension of  IceCube plus surface array plus radio~detector\\
                                                  
\midrule
KM3NeT Phase\,3  &               & Mediterr.\ Sea   & ca. 3    & Possible extension of  ARCA to 6~blocks \\
   
\bottomrule                                                        
\end{tabularx}
\end{table}

P-ONE ({\it Pacific Ocean Neutrino Experiment})\,\cite{P-ONE:2020ljt} is a 
project in its R\&D and prototype phase, envisaging a neutrino telescope in the Pacific Ocean
off the coast of British Columbia, Canada. It will strongly profit from an existing deep-sea cable
infrastructure. 

Three Chinese collaborations are currently working towards projects on the
10\,km$^3$ scale. TRIDENT ({\it Tropical Deep-Sea Neutrino Telescope}) is 
planned to be installed in the South China Sea, with an instrumented volume
of 7.5\,km$^3$. R\&D and site exploration are ongoing. A pathfinder array
(TRIDENT Phase-1) is planned to be deployed in 2026/2027 \cite{TRIDENT:2022hql}.
HUNT ({\it High-Energy Underwater Neutrino Telescope}) envisions an instrumented
volume of 30\,km$^3$, either in the South China Sea or Lake
Baikal \cite{Huang:2023mzt}. NEON ({\it NEutrino Observatory in the Nahnhai})
aims at another detector of 10\,km$^3$ in the South China
Sea \cite{Zhang:2024slv}. Whether there will be some convergence among
these three projects appears open at present.

The IceCube-Gen2 project \cite{IceCube-Gen2:2020qha} is planned to extend the sensitivity
of IceCube towards higher energies.  It would consist of an 8\,km$^3$ optical array with IceCube at its center, combined
with a radio array (see below) for highest-energy neutrinos. A surface array would provide
a veto against atmospheric events.

\subsection{Detectors for the Tens of PeV to EeV Range}
\label{PeV}

The technology of optical Cherenkov detectors reaches its limits for energies of a few tens of PeV. The neutrino flux decreases steeply with energy, requiring instrumented areas on the 100 km$^2$ scale. The range of Cherenkov light in water and ice, however, limits the maximum distance between photomultipliers to $\sim $200\,m, restricting the scalability of this technology to sizes of a few 10\,km$^3$ at best.

At energies above 100 PeV, large air shower detectors like the Pierre Auger Observatory in Argentina 
\cite{PierreAuger:2025was} or the Telescope Array in Utah, USA 
\cite{TelescopeArray:2019mzl}, are seeking horizontal air showers induced by neutrino interactions {\it deep} in the atmosphere (showers induced by charged cosmic rays start {\it at the top} of the atmosphere and lead to different spatial and time patterns than those induced by neutrinos). This method reaches its optimum sensitivity in the multi-EeV range. A slightly lower threshold is obtained for tau neutrinos that skim Earth and interact close to the array. The $\tau$ lepton produced in the interaction can escape the rock and decay into hadrons. The resulting particle cascade would be recorded by the surface detector or the fluorescence telescopes.

Another path to highest energies is detecting the radio waves emitted by high-energy tightly focused particle showers. The method is based on the Askaryan effect \cite{Askaryan:1990wf}:  electromagnetic showers develop a negative charge excess of electrons along their path. This net charge propagates like a disc that is 1\,cm thick and 10\,cm in diameter. Each particle emits Cherenkov radiation and, since the distance between particles is smaller than radio wavelengths, coherence occurs and the signal strength is proportional to the {\it square} of the number of electrons in the shower, which in turn is proportional to the shower energy. As a result, the strength of the Askaryan radiation increases as the square of the energy of the original~particle. 

One realization of this principle is to install radio antennas in ice. The attenuation length of radio waves in ice is about a kilometer, which is why radio detectors can be placed in the ice at ten times the distance of light detectors. Over the past decades, various aspects of the technology have been investigated in Antarctica
\cite{ARA:2024rfo,Anker:2020lre}. A medium-sized facility is currently under construction in the 3\,km thick ice sheet close to the US Summit station in Greenland: RNO-G (the {\it Radio Neutrino Observatory Greenland})  \cite{RNO-G:2024esr}. It will consist of about 30 antenna stations, to be installed on a one-kilometer grid, covering a total area of about 25\,km$^2$. RNO-G is blind to energies below a few tens of PeV, but, beyond this energy, its sensitivity grows steeply. RNO-G is intended to measure neutrinos in the 100 PeV range and will also serve as a test setup for a future much larger antenna array at the South Pole (as part of IceCube-Gen2).

Another realization to detect ultra-high-energy neutrinos via radio signals is to use mountains as natural neutrino targets. Cosmic tau neutrinos would interact in a mountain producing $\tau$ leptons.  At 100 PeV, the average decay length of $\tau$ leptons is about 50\,m. The decay can take place, for example, in a valley behind the mountain. It leads to a tightly focused particle shower, which emits radio signals in the MHz and GHz range. The signal can be recorded by radio antennas on the opposite mountain slope. Several projects of this type are currently under development, led by {\it GRAND}, the Giant Radio Array for Neutrino Detection in western China \cite{GRAND:2025jjc}, and to be followed by others in South America and Africa. Other detector projects envision balloon- or space-based detectors. See  \cite{Coleman:2022abf} for a review of these future detectors and the PDG review of S.\,Klein and A.\,Nelles in \cite{ParticleDataGroup:2024cfk}.

\vspace{3mm}
\section{The Diffuse Flux of Cosmic Neutrinos}
\label{diffuse-section}

\begin{figure}[H] 
\includegraphics[width=.98\textwidth]{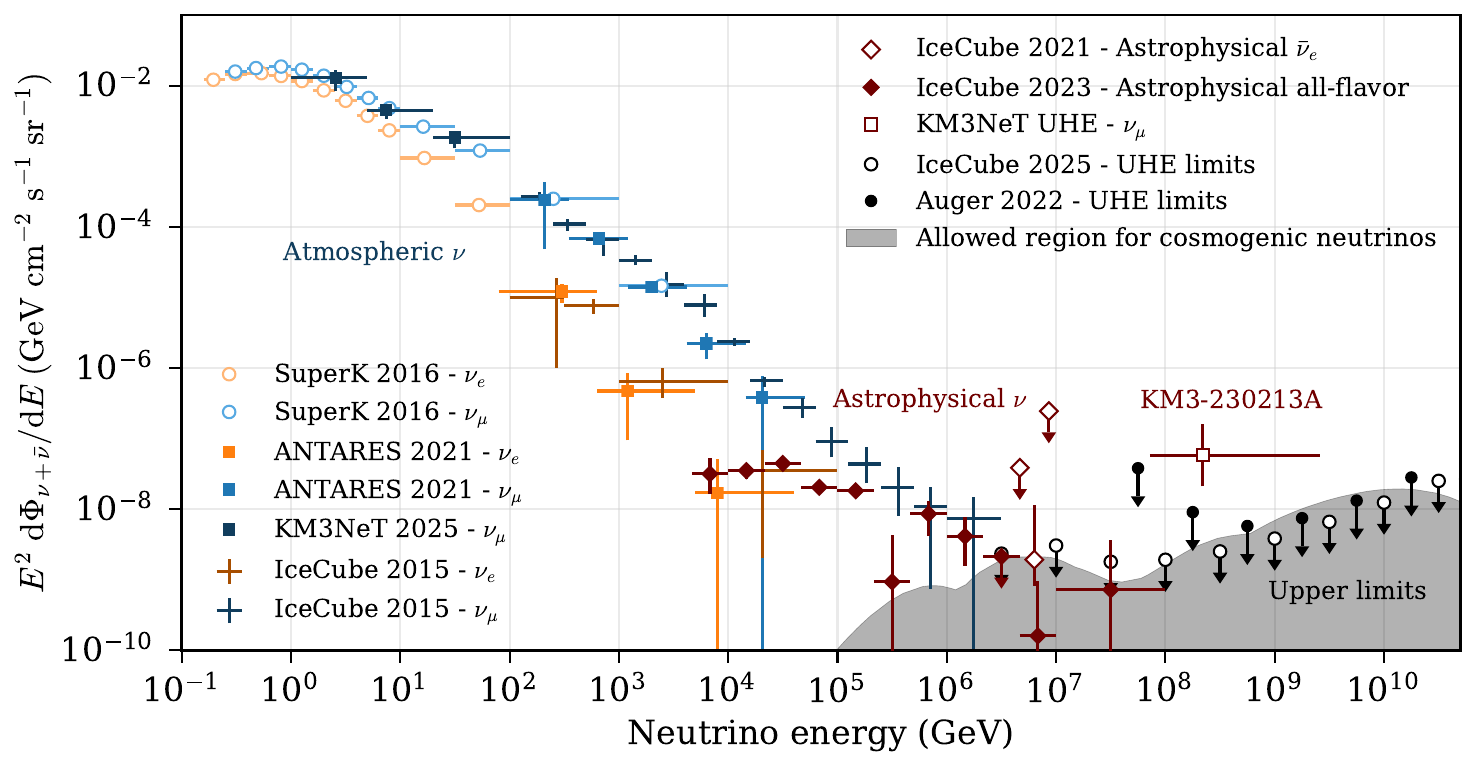}
\caption{Measured 
 energy spectra of atmospheric and cosmic diffuse neutrino fluxes. Experimental limits at highest energy are compared to model predictions for cosmogenic neutrinos  \cite{Abdul_Halim_2023}.
All the fluxes are normalized to one flavor ($\nu+\overline\nu$) assuming a ratio $\overline\nu  : \nu=1$ and, for cosmic neutrinos, a ratio $\nu_e : \nu_{\mu} :\nu_{\tau} $ = 1:1:1 at Earth. 
Data sources:  
SuperK 2016 \cite{Super-Kamiokande:2015qek};
ANTARES 2021 \cite{ANTARES:2013iuz}; 
KM3NeT 2025 $\nu_\mu$ \cite{KM3NeT:2025pne}; 
IceCube 2015 $\nu_e$ \cite{IceCube:2015mgt};
IceCube 2015 $\nu_{\mu}$ \cite{Borner:2015sed};
IceCube 2021 astrophysical $\bar\nu_e$ ($\nu+\overline\nu$ flux derived from a single candidate event for the Glashow resonance) \cite{IceCube:2021rpz};      
IceCube 2023 astrophysical all-flavor~\cite{IceCube:2023qpn};  
KM3NeT-UHE $\nu_\mu$ \cite{KM3NeT:2025npi}.  
The limits at the highest energies are taken from \cite{IceCubeCollaborationSS:2025jbi} (IceCube) and \cite{PierreAuger:2023pjg} (Auger). They are shown in the energy bins used for the corresponding analyses and have been adjusted to bin widths of one decade. The region allowed for the expected cosmogenic neutrino fluxes are taken from \cite{vanVliet:2019nse} (Figure taken from U. Katz and C. Spiering in \cite{ParticleDataGroup:2024cfk}).
}
\label{diffuse_flux_total}
\end{figure}

The diffuse flux of neutrinos with energies larger than $\sim$100 MeV  is composed of atmospheric neutrinos, dominating the spectrum up to a few hundred TeV, and of astrophysical (or ``cosmic'') neutrinos from distant sources. 
 Figure~\ref{diffuse_flux_total} shows a compilation of measurements of the diffuse flux of energetic neutrinos.

The flux of atmospheric neutrinos  has been measured by many experiments. The production chain for {\it conventional} atmospheric neutrinos follows Equations\,(\ref{CR-interactions})--(\ref{piminusdecay}), with an observed ratio $\nu_e:\nu_\mu \approx 1:2$ at energies of a few GeV (where almost all the muons decay in the atmosphere) and a ratio $<$\,1:10 
 at TeV energies, where the muons reach the ground before decaying.  The flux of upward-moving muon neutrinos with energies $<$\,100\,GeV is reduced due to neutrino oscillations on their way through Earth, lowering the $\nu_e:\nu_\mu$ ratio at the smallest energies. 

The flux of conventional atmospheric neutrinos follows an $E_{\nu}^{-3.7}$ power law.  On top of this dominant component, there must also be a contribution from {\it prompt} atmospheric neutrinos produced by the decay of hadrons containing a charm or bottom quark. Since these heavy hadrons are produced early in the air shower and decay quickly before losing energy, the energy spectrum of prompt neutrinos must be harder ($\sim$\,$E_{\nu}^{-2.7}$) than that of atmospheric neutrinos. This component is expected to dominate the atmospheric neutrino flux at its highest energies. Predictions for the prompt neutrino flux have large uncertainties related to energy spectrum and mass composition of cosmic rays, the model used to describe heavy-flavor production, and the used parton distribution functions 
\cite{PhysRevD.107.023014}. Present measurements of the neutrino spectrum provide just upper limits for the prompt~flux. 

Atmospheric neutrinos are of interest to particle physics --- in particular oscillation physics (see, e.g., \cite{Jung:2001dh,Brunner:2019nqy,IceCube:2017lak}) --- and Earth science \cite{Donini:2018tsg,IceCube:2025utw}.

\vspace{1mm}
\subsection{Size and Shape of the Diffuse Flux of Cosmic Neutrinos}\label{sec4.1}

The identification of a diffuse flux of cosmic neutrinos rests mainly on an excess at high energies: the flux of atmospheric neutrinos behaves like $E^{-3.7}$ and that of cosmic neutrinos like $E^{-2.0 ... 2.5}$.  Indications for such an excess were already found using data taken with the 40-string and 59-string configurations of IceCube \cite{PhysRevD.89.102001,IceCube:2013gge}, with 2.7\,$\sigma$ and  1.8$\,\sigma$ excesses above the background expectation, respectively. Simultaneously, the first two PeV neutrinos were found
\cite{IceCube:2013cdw}. Shortly later, data from the 79-string configuration and the first two years of the final configuration with 86 strings were analyzed, resulting in a 4.1\,$\sigma$ significance for an excess over the flux of atmospheric neutrinos \cite{IceCube:2015vkp}. The upper limits from the first two analyses were consistent with each other, while their best-fit values were consistent with those of the mentioned three-year analysis. 

The success of the mentioned three-year analysis was not only due to the larger statistics but also the application of the so-called HESE (high-energy starting event) method, which had been proposed a few years before \cite{PhysRevD.79.043009} and applied here for the first time.
The method aims to identify cosmic neutrinos by rejecting not only the background of atmospheric muons but also reducing that of atmospheric neutrinos. It selects only events with an interaction vertex contained in the detector (“starting events”). Moreover, outer parts of the detector are used as a veto layer. Atmospheric neutrinos from air showers above the detector are typically accompanied by downgoing muons from the same shower. These muons enter the detector from above and activate the veto.  

\begin{figure}
\includegraphics[scale=0.9]{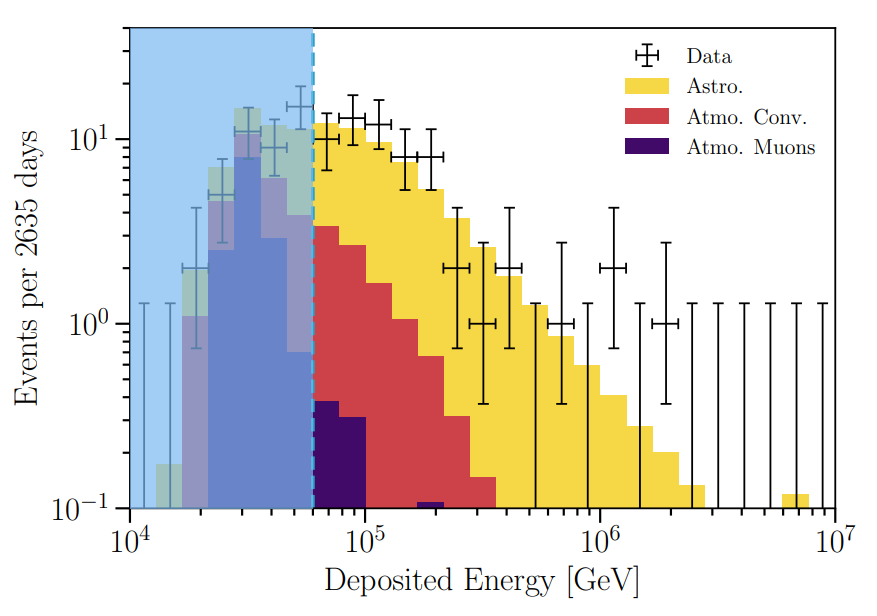}
\caption{
Measured (data points) and expected (histograms) distributions of the deposited energy measured in the IceCube detector for events passing the HESE selection, which strongly suppresses downgoing atmospheric neutrinos by vetoing accompanying muons. The cosmic neutrino flux has been assumed to follow an unbroken power law in $E_\nu$. Its normalization and spectral index, as well as the normalizations of the conventional and prompt atmospheric neutrino and the atmospheric muon contributions, were fitted to the data, where the region below $60\,$TeV (blue) was excluded from constraining the cosmic neutrino flux. Figure taken from
\cite{PhysRevD.104.022002}.} 
\label{cosmicray:fig:nu-extragalactic}      
\end{figure}

Figure~\ref{cosmicray:fig:nu-extragalactic} shows the result for the 7.5-year HESE sample \cite{PhysRevD.104.022002} with its clear excess of cosmic neutrinos at high energies.
The zenith angle distribution (not shown) of events appears to be almost flat as a function of cosine zenith. A small decline towards the upgoing region is due to Earth’s absorption of high-energy neutrinos. The flat behavior for downgoing events convincingly demonstrates the strong contribution of cosmic neutrinos. In contrast to those, a large part of downgoing {\it atmospheric neutrinos} are vetoed by the HESE trigger. Downgoing atmospheric neutrinos are therefore strongly suppressed.

\begin{figure}
\includegraphics[scale=0.75]{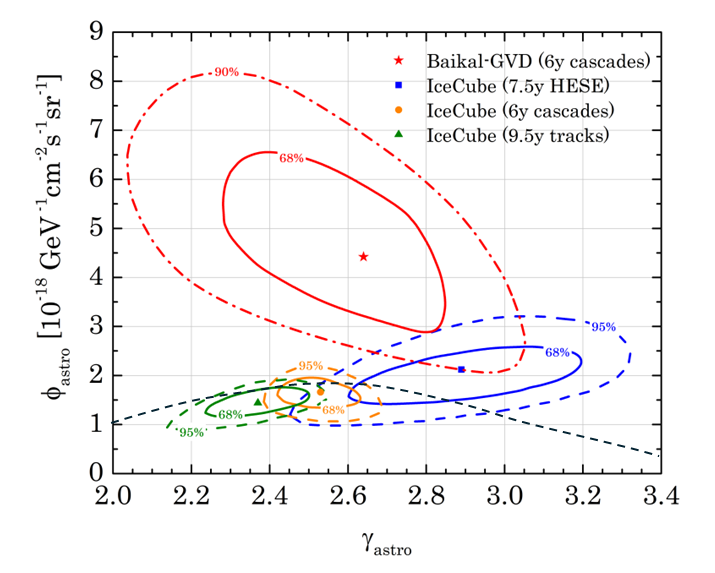}
\caption{The best$-$fit
parameters and the contours at 68\% and 90\% (Baikal-GVD) and 95\% (IceCube) confidence levels for a single power-law hypothesis according to Equation (\ref{flux-formula}). The figure has been taken from \cite{GVD:2025lya}. The black dashed curve has been added by hand (taken from \cite{ANTARES:2024ihw}) and defines their 95\% posterior probability credible area. The IceCube data refer to analyses of high-energy starting events (blue \cite{PhysRevD.104.022002}), track-like events (green \cite{IceCube:2021uhz}) and cascade-like events 
(orange \cite{IceCube:2020acn}).
}
\label{simple_power_law}      
\end{figure}

In the following, high-energy excesses were also identified for tracks entering the detector from outside \cite{IceCube:2021uhz} and fully contained cascade events \cite{IceCube:2020acn}.  Fitting the IceCube data to
\begin{equation}
\Phi = \Phi_{0} \left(\frac{E_{\nu}}{E_0}\right)^{-{\gamma}},
\label{flux-formula}
\end{equation}

\noindent
the normalization factor for all the analyses is about
$\Phi_{0} \approx 10^{-18}$\,GeV$^{-1}$\,cm$^{-2}$\,s$^{-1}$\,sr$^{-1}$, while the spectral index $\gamma$ varies between 2.0 and 2.5; see Figure~\ref{simple_power_law}.

Some indications for a diffuse cosmic flux have also been reported by ANTARES in 2020 (see, for the latest summary, \cite{ANTARES:2024ihw}), albeit with much lower significance (resulting in the 95\% confidence region shown in  Figure~\ref{simple_power_law}). Baikal-GVD also observes an astrophysical diffuse flux of neutrinos at a significance of 5.1\,$\sigma$ \cite{GVD:2025lya} but with a substantially higher normalization factor than those from the various IceCube analyses. Judging the compatibility/incompatibility of the individual analyses, one has to take into account that they cover different energy regions. ANTARES results refer to energies below 50\,TeV, while HESE results (IceCube) and the results from Baikal-GVD are mainly based on data above 100 TeV. A possible reason for the high $\Phi_0$ value of GVD could be an underestimation of the effective area in Monte Carlo simulations. Whatever the reason for the disagreement, the situation clearly calls for data from other experiments, with KM3NeT being that which could yield it in the next years.

\unskip

\begin{figure}
\includegraphics[scale=0.83]{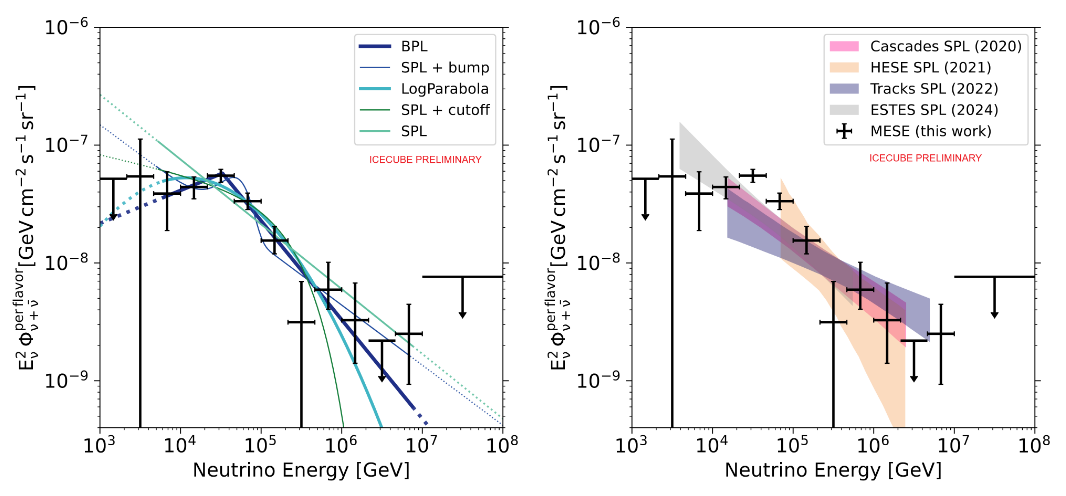}
\caption{Results of 
 a fit of the astrophysical neutrino flux in independent energy
bands (IceCube). Left: The results are compared to the models fitted in the analysis. The solid lines show the energy range where the dataset is sensitive to the respective model, and the dotted lines show the energy range over which the fit is performed. Right: The segmented fit is compared to previous measurements from IceCube, all favoring the SPL model. The shaded region corresponds to the 68\% confidence region for these measurements, as do the error bars on the MESE points. MESE stands for medium-energy starting event (see text), ESTES for enhanced starting track selection, SPL for single power law, and BPL for broken power law.
Figure taken from \cite{IceCube:2025dlr}.
}
\label{segmented-fit}      
\end{figure}

The energy-dependent variations in the value for the spectral index $\gamma$ in Figure \ref{simple_power_law} indicate that a simple power law is likely too simple a description of the spectrum. This has been investigated in two recent papers \cite{IceCube:2025tgp, IceCube:2025dlr}. Figure~\ref{segmented-fit} shows a clear steepening
of the spectrum above $\sim$30 TeV. The analysis is based on the medium-energy starting event (MESE) selection. It applies a series of vetoes to separate a sample enriched with cosmic neutrinos from 1 TeV to 10 PeV, covering the entire sky. Data have been fitted with different spectrum hypotheses.  A broken
power-law fit yields the best agreement with the data (Figure \ref{segmented-fit} (left)) and is preferred over the simple power-law option with a significance of 4.7\,$\sigma$. The fitted spectral indices below and above the break are $\gamma_1 = 1.7 \pm 0.3$ and $\gamma_2 = 2.8 \pm 0.1$, respectively.

The maximally expectable flux (or the energy density) for diffuse cosmic neutrinos has long been related to the flux of gamma rays in the MeV--GeV region \cite{Berezinsky:1969erk} and the flux of cosmic rays \cite{Waxman:1998yy}. The latter bound, the ``Waxman--Bahcall upper bound'',  was a final confirmation that a cubic-kilometer scale detector would be necessary to detect the diffuse flux of cosmic neutrinos.

Indeed, the energy density $E^2\Phi$ of the diffuse neutrino flux has been compared to that of gamma rays and cosmic rays (see, e.g., \cite{Ahlers:2018dtq,Fang:2022trf}). It turns out to 
match that of the isotropic gamma-ray background as measured with the Fermi satellite and that of cosmic rays above $10^{19}$ eV, as observed by the Pierre Auger Observatory and TALE.  The authors of~\cite{Fang:2022trf} noted that (for optically thin sources) the measured flux of TeV and sub-TeV neutrinos would require a higher flux of gamma rays than Fermi has measured in the region of 1--100 GeV. The result therefore indicates that the observed neutrinos have been produced in environments that are optically thick to gamma rays with energies above 1~GeV.
This nicely confirms the concept of ``cocooned'' sources proposed many years ago by V. Berezinsky (see, e.g., \cite{Berezinsky:1980mh}).

One should note that improved extrapolations towards lower energies using, e.g., segmented fits like that of \cite{IceCube:2025dlr} may have a strong impact on the relation between Fermi and neutrino data. The smaller spectral slope towards lower neutrino energies obtained in \cite{IceCube:2025dlr} weakens the conclusions of \cite{Fang:2022trf}.  It remains to be seen what further improved IceCube data and KM3NeT data will reveal (the Baikal GVD data may have too high a threshold to contribute here). In any case, this question nicely demonstrates the concept of multimessenger astronomy, even on the level of diffuse fluxes.

\subsection{Flavor Composition of the Diffuse Flux of Cosmic Neutrinos}

In most astrophysical scenarios, neutrinos are produced through the
$\pi/K\to\mu\to e$ decay chain, resulting in a flavor ratio
$\nu_e$:$\nu_\mu$:$\nu_\tau\approx $ 1:2:0. If the emission region is dense, muons can lose most of their energy before decaying, resulting in a ratio 0:1:0 in the extreme case. In a third scenario, protons are kept in the source region by magnetic fields. 
Only neutrons would escape and then decay via $n \to p + e^- + \overline\nu_e$, resulting in a ratio 1:0:0. Over cosmic distances, flavor mixing turns these ratios to 0.30:0.36:0.34 for the first source model (1:2:0), 0.17:0.45:0.37 for the second (0:1:0) and  0.55:0.17:0.28 for the third (1:0:0) \cite{IceCube:2020fpi}. 

\begin{figure}[H]

\includegraphics[scale=0.8]{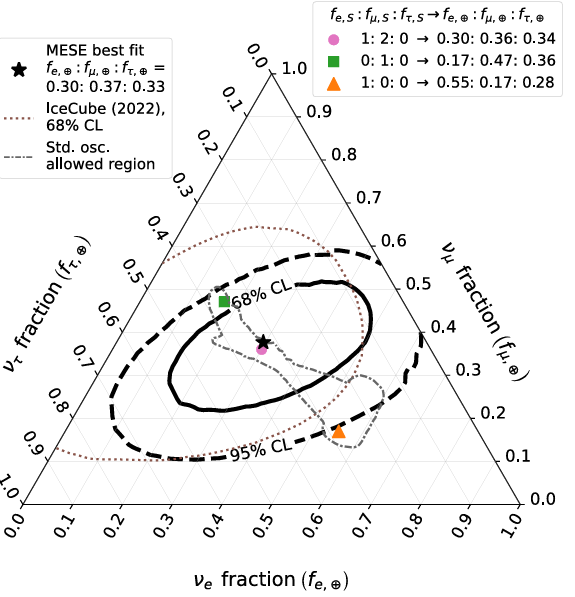}
\caption{
Ternary diagram of the results of the flavor
composition fit: \hl{} 
 The axes show the fraction of $\nu_e$, $\nu_{\mu}$ and $\nu_{\tau}$
at Earth. 68\% and 95\% CL contours are shown as solid and
dashed lines, respectively. Expected flavor composition at
Earth after standard oscillations for benchmark production
mechanisms (pion decay: circle; muon damping: square; neutron decay: triangle) and all possible flavor compositions after
propagation (dash--dot line) are shown. The dotted
line shows the 68\% CL contour from a previous IceCube measurement. Figure taken from \cite{IceCube:2025ole}.
}
\label{flavor}      
\end{figure}

The flavor ratios can be determined according to the event signatures sketched in Figure\,\ref{signatures}. Track events are mostly due to charged current (CC) interactions of $\nu_{\mu}$, cascade events to CC interactions of $\nu_{e}$ and to NC reactions of all types of neutrinos. CC interactions of some  $\nu_{\tau}$ appear as track events if the decay length of the $\tau$ is small and double-bang events if it is large enough to generate a double-cascade pattern.
Even if the decay length is too small to generate a clear double bang pattern, $\tau$ decays can be identified from the double-pulse pattern recorded by individual photomultipliers.
The first tau neutrino candidates in IceCube data were identified in 2020
\cite{IceCube:2020fpi}.
A recent IceCube study of event morphologies 
\cite{IceCube:2025uyt}
slightly favors $\nu_e$:$\nu_{\mu}$:$\nu_{\tau}$ = 1:2:0 at the source (pion decay according to Equations
(\ref{piplus-decay}) and (\ref{piminusdecay}))
versus 0:1:0 (damped muons) and disfavors a 1:0:0 scenario (neutron decay). In \cite{IceCube:2025ole} (based on MESE data, with 4960 cascade events, 4919 track events and 9~double cascade events), the best fit for the broken power law yields
a ratio 0.30:0.37:0.33 
(see Figure~\ref{flavor}).

\vspace{0.5cm}

\subsection{Two Peculiar Events}\label{sec4.3}


\textls[-35]{{\bf The ``Glashow'' event}: 
 IceCube has recorded a single event with energy \mbox{$6.05\pm0.72$ PeV~}\cite{IceCube:2021rpz}}. It is compatible with having been generated by the resonant process $\bar\nu_e + e^- \to W^-\to\text{hadrons}$ at $E_\nu=6.3$ PeV (Glashow resonance) \cite{Glashow:1960zz}. The event is highlighted in Figure \ref{diffuse_flux_total} as  ``IceCube 2021 Astrophysical $\bar{\nu}_e$''.

A statistically significant observation of the Glashow resonance would prove the presence of $\bar\nu_e$  in the recorded flux. Since the ratio $\bar\nu_e$:$\nu_e$ depends on the mass composition of cosmic rays and also the photon density and the magnetic field strength of the source~\cite{Biehl:2016psj}, this could provide information about the environment of the production process.


{\bf The 220-PeV event KM3-230213A}: On 23 February 2025, KM3NeT announced the observation of a neutrino event with an estimated neutrino energy of $220^{+570}_{-110}$\,PeV in the ARCA detector \cite{KM3NeT:2025npi}. This is the neutrino with the highest energy ever recorded, an order of magnitude above the most energetic neutrinos detected with IceCube. The diffuse neutrino flux corresponding to that event is indicated in Figure~\ref{diffuse_flux_total} as ``KM3-230213A''. 
Interpretations of the event include a cosmogenic or even galactic origin, as well as a possible association to blazars (see \cite{Coniglione:2025fpz} and references therein), but no clear indication of any of these hypotheses has been found.  

\begin{figure}[H]
\includegraphics[scale=0.65]{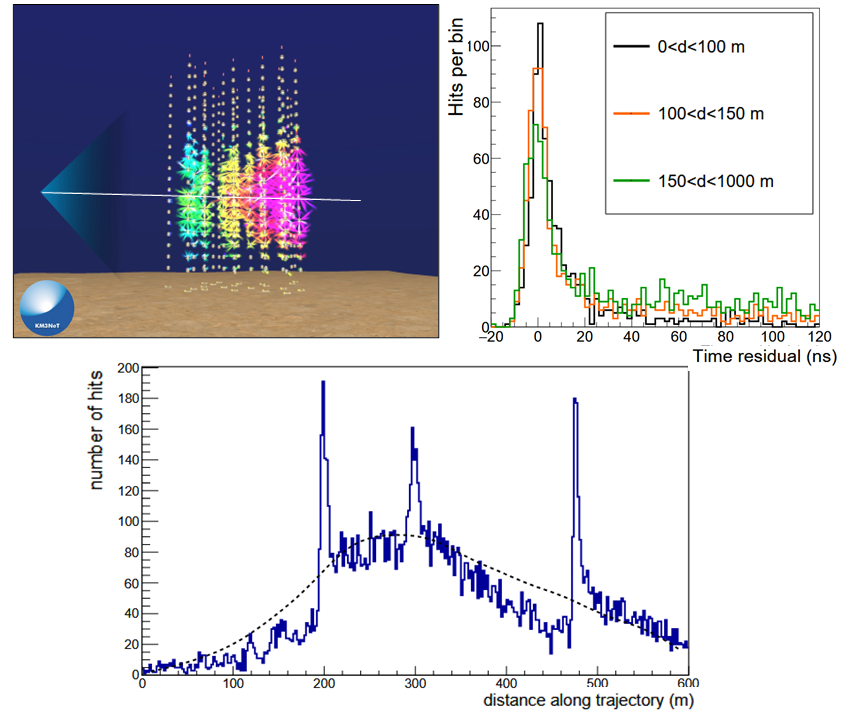}
\caption{
\textbf{Top left}: 
Side view of KM3-230213A. The reconstructed muon trajectory, from right to left, is
shown as a white line. 
The hits on individual PMTs are represented by cones along the axis of the PMT  orientation, with the height of the cone indicating the number of detected photons and the color indicating the detection time (purple: early; cyan: late). \textbf{Top right}: Distribution of the time
residual assuming a single muon using the first hits on each PMT. The different lines correspond to slices in radial distances from the track. \textbf{Bottom}: Distribution of the estimated photon emission point along the reconstructed muon trajectory,
represented as the distance to the starting point of the track (when entering the detector). The dashed line shows the average expectation for a 100 PeV muon with the same direction and position as KM3-230213A \cite{KM3NeT:2025npi}.}
\label{220PeV}
\end{figure}

Figure \ref{220PeV} displays some illustrative data on that event.  Note the distinct appearance of three cascades from high-energy bremsstrahlung or nuclear interactions along the trajectory (bottom). This and the distribution of the time residuals (top right) demonstrate the superior capability for pattern recognition, which distinguishes water detectors from ice detectors with their much stronger light scattering effects. 

\subsection{Cosmogenic Neutrinos}  \label{sec4.4}
Cosmogenic neutrinos are decay products of charged pions produced in interactions of ultra-high-energy cosmic rays with the cosmic microwave (a) or infrared (b) background radiation ($p + \gamma \to n + \pi^+$) \cite{Berezinsky:1969erk}. They can also originate in the decay of neutrons produced in photodisintegration processes \cite{Ave:2000nd,Hooper:2004jc} (c).  The neutrino flux at EeV energies is expected to be dominated by (a) and (c).

The expected flux can be estimated from the measured spectrum and composition of charged cosmic rays, the cosmic microwave density and the properties of proton and nuclear interactions with photons at center-of-mass energies in the GeV range. Current uncertainties of these estimates are one order of magnitude or even more. Figure~\ref{diffuse_flux_total} shows the presently tightest limits, together with the shadowed region for cosmogenic neutrino fluxes that are still consistent with those limits (status 2019) and cosmic-ray analyses \cite{vanVliet:2019nse}.

\begin{figure}[H]
\includegraphics[scale=0.68]{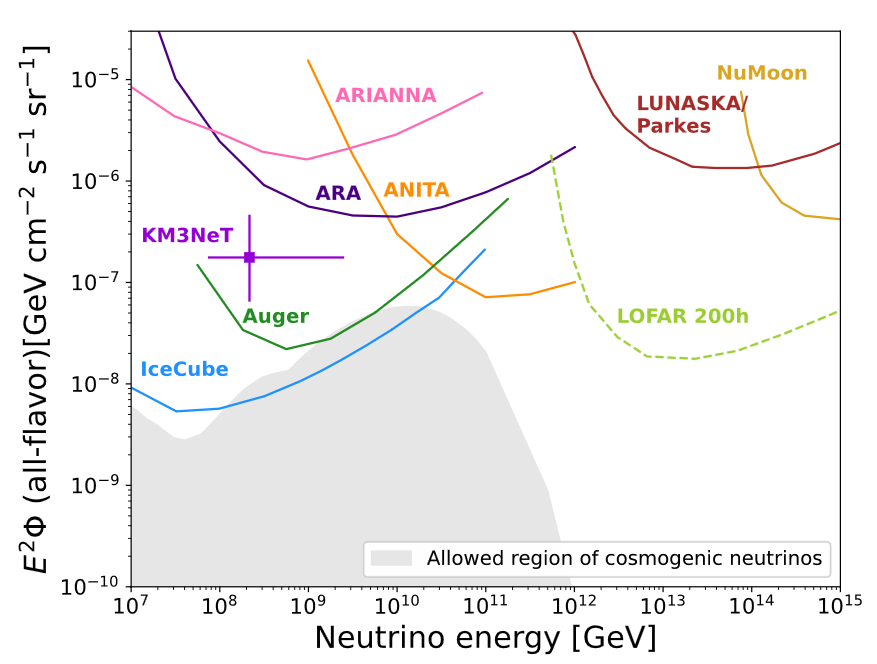}
\caption{
Representative 3-flavor (summed assuming equal fluxes of each flavor) differential (over one decade in energy) limits from different experiments and prototype experiments. Shown are limits from the IceCube ultra-high-energy neutrino search, from the Auger search for earth-skimming $\nu_{\tau}$, from the LUNASKA/Parkes and NuMoon lunar searches, and from ANITA, ARA and ARIANNA, searching for neutrino-initiated radio emission in Antarctic ice. Also shown is a projection for LOFAR and the flux implied by the 220-PeV event seen by KM3NeT.  The shaded area is the allowed region for cosmogenic neutrinos from a recent global analysis as published in \cite{b:2025bzp}. Figure taken from \cite{ParticleDataGroup:2024cfk}.
}
\label{cosmogenic}      
\end{figure}

Figure~\ref{cosmogenic} shows the ultra-high-energy region and some limits in more detail. Expected limits from future experiments are not included.  LOFAR is already operating but did not yet publish neutrino results, so the dashed LOFAR curve is only a projection.
The KM3NeT event KM3-230213A with its most probable energy of 220 PeV is in tension (2.9\,$\sigma$) with the shown IceCube flux limit.  The  LUNASKA/Parkes and NuMoon experiments have searched for radio emission from neutrinos interacting in the lunar regolith, the ANITA balloon experiment and the ARA and ARIANNA prototype experiments for neutrino-initiated radio emission in Antarctic ice. ARA and ARIANNA are prototype devices, so their limits should just be taken to indicate the energy range to be covered, with much higher sensitivity, by larger arrays like RNO-G, the Radio Neutrino Observatory in Greenland, or the 500 km$^2$ radio extension of the planned IceCube-Gen2 observatory at the South Pole.

\section{Neutrinos from the Galactic Plane}
\label{GP-section}

 Although the flux of high-energy cosmic neutrinos is largely extragalactic, 
 a galactic component is inevitably produced when cosmic rays kept magnetically in the galaxy interact with interstellar gas.
 These cosmic rays constitute a non-thermal component of the Milky Way with a comparable energy density to that of magnetic fields and starlight. Their interactions with the interstellar gas therefore represent a significant channel of energy dissipation within the interstellar medium. The decay of secondary hadrons produced in these collisions will lead to a diffuse flux of neutrinos from the Galactic Plane.
 
 Various models have been developed to calculate the flux and the spatial distribution of these neutrinos. The most obvious is based on Fermi-LAT observations of diffuse gamma rays in the GeV range. Assuming a certain power law in energy, these data can be extrapolated to the TeV range.  
 If the gamma rays stem from decays of $\pi^0$ co-produced with
 $\pi^{\pm}$, one can calculate fluxes of neutrinos from  $\pi^{\pm}$ decays.
 Other models start with modeling the propagation of charged cosmic rays in the galaxy and their interactions with galactic dust. Both approaches yield an intensity distribution of high-energy neutrinos across the Milky Way.
 
\begin{figure}[H]

\includegraphics[scale=0.80]{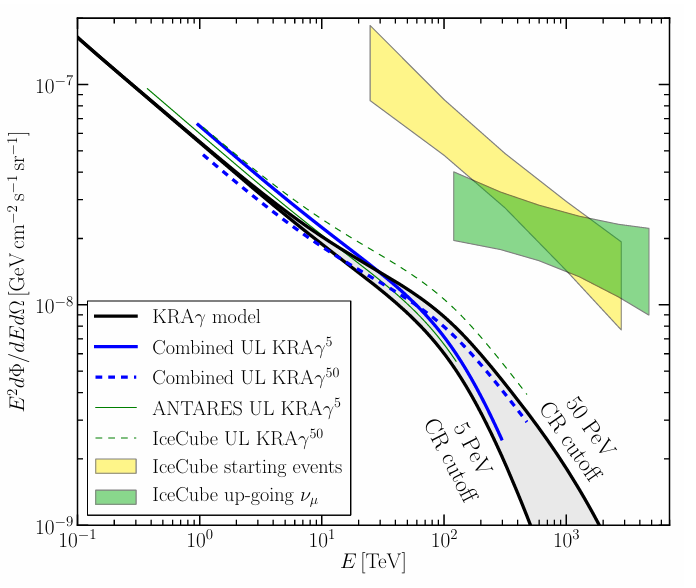}
\caption{Combined 
 upper limits (ULs) at 90\% confidence
level (blue lines) on the three-flavor neutrino flux of the
KRA$_{\gamma}$ model with the 5 and 50 PeV cutoffs (black lines).
The boxes represent the diffuse astrophysical neutrino fluxes
measured by IceCube using an isotropic flux template with
starting events (yellow) and upgoing tracks (green).
Figure taken from \cite{ANTARES:2018nyb}. \copyright{AAS}. Reproduced with permission.
}
\label{galactic-models}      
\end{figure}

In 2018, the ANTARES and IceCube collaborations published a combined search for neutrinos from the Galactic Plane using ten years of ANTARES track and cascade and seven years of IceCube track events. The data have been combined into a joint likelihood test for neutrino emission according to a model developed in \cite{Gaggero:2015xza}. One of the free parameters of this so-called KRA model is the cutoff of charged galactic cosmic rays in the PeV region, e.g., 5 PeV per nucleon for KRA$\gamma_5$ and 50 PeV for KRA$\gamma_{50}$.
Figure \ref{galactic-models} shows the results. 
No significant excess was found. The limits obtained, however, started constraining the
model parameter space for cosmic ray production and transport in the galaxy. Note also that neutrinos from the Galactic Plane can contribute only a very small part of the isotropic cosmic neutrino flux.

In 2022, ANTARES developed an analysis focusing on the Galactic Ridge and obtained a $1.3\,\sigma$ hint for TeV emission from an ``on-zone'' region of galactic longitude $\lvert l \rvert < 30^{o}$ and galactic latitude $\lvert b \rvert < 2^{o}$ \cite{ANTARES:2022izu}. 
In 2025, a final all-flavor neutrino dataset, collected over 15~years by ANTARES, was analyzed \cite{ANTARES:2025wvi} and yielded upper limits that are compatible with the galactic flux measured by IceCube in 2023 and described in the following paragraphs.

\begin{figure}[H]
\includegraphics[scale=0.38]{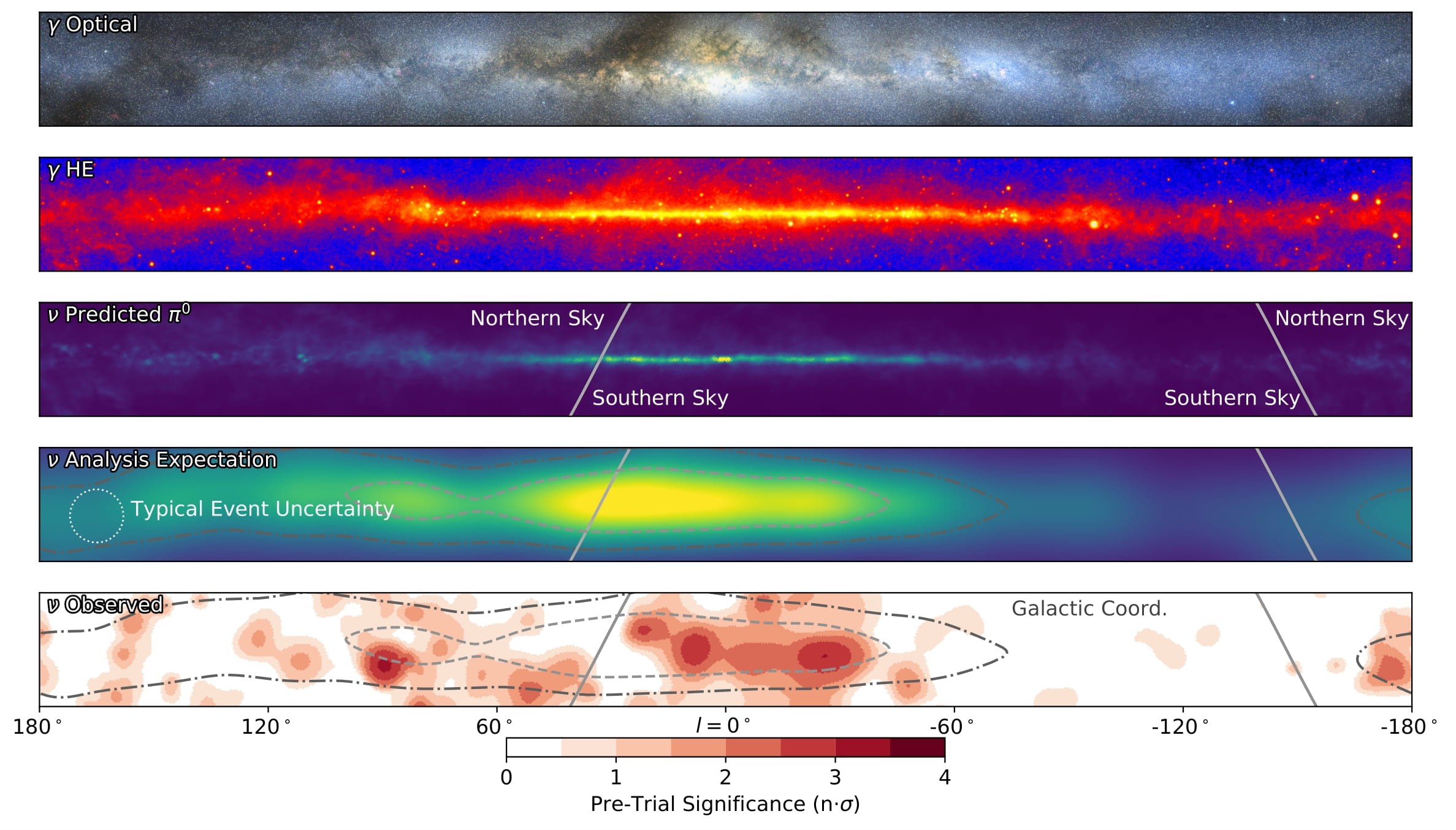}
\caption{
The plane 
 of the Milky Way galaxy in photons and neutrinos. Each panel --- from A (top) to E (bottom) --- is in
galactic coordinates, with the origin being at the Galactic Center,
extending to $\pm15^{o}$ in latitude and $\pm180^{o}$ in longitude. (\textbf{A}) Optical color image, which is partly obscured by clouds of gas and dust that absorb optical photons. (\textbf{B})
Integrated flux in gamma rays from the Fermi-LAT 12$-$year
survey at energies greater than 1\,GeV. (\textbf{C}) Emission template calculated for the
expected neutrino flux, derived from the $\pi^0$ template that matches the Fermi-LAT observations
of the diffuse gamma-ray emission. (\textbf{D})~The emission template from panel (\textbf{C}) including
the detector sensitivity to cascade-like neutrino events and the angular uncertainty of a typical
signal event ($7^{o}$, indicated by the dotted white circle). Contours indicate the central regions
that contain 20\% (50\%) of the predicted diffuse neutrino emission signal. (\textbf{E})  Pre-trial
significance of the IceCube neutrino observations, calculated from all-sky scan for point-like
sources using the cascade neutrino event sample. Contours are the same as panel (\textbf{D}). Gray
lines in (\textbf{C}--\textbf{E}) indicate the northern--southern sky horizon line at the IceCube detector.
Figure taken from \cite{IceCube:2023ame}; see there also for references.
}
\label{galactic-KI}      
\end{figure}

The breakthrough came with an IceCube analysis published in 2023 \cite{IceCube:2023ame}. It was based on a novel event selection that used machine learning techniques to select cascade events with much higher efficiency. It was applied to ten years of IceCube data.
For each of the mentioned three models, templates of the expected distribution
for directions and energies of neutrinos were derived, taking
into account the trigger efficiencies and the reconstruction accuracy for the selected IceCube cascade events. The observed distributions in energy and galactic coordinates were compared with the expected distributions.  Figure~\ref{galactic-KI} demonstrates the steps for a likelihood function based on the gamma-ray data measured by Fermi according to the $\pi^0$ connection explained above.

The central parts of the galaxy are in the Southern Hemisphere. Therefore, the inclusion of throughgoing tracks was not possible because that sample would have been dominated by downgoing atmospheric muons. Contained cascade events as used in this analysis, however, suffer from an inferior angular resolution compared to track events. This explains the strong smearing when going from the third to the fourth panel in the figure. The comparison of the three diffuse emission models to a background-only hypothesis excluded the latter, with statistical significances of $4.71\,\sigma$,  $4.37\,\sigma$ and $3.96\,\sigma$ for the $\pi^{o}$, KRA$\gamma_5$ and KRA$\gamma_{50}$ templates, respectively. Taking into account the trial factor, the final significance is $4.5\,\sigma$. While the signal is consistent with models for the diffuse emission from the Galactic Plane, unresolved point sources  will also contribute to some extent.
An updated analysis with additional 2.5 years of data that also incorporates track events increases the sensitivity such that a detection with a significance clearly
above 5$\,\sigma$ is expected \cite{IceCube:2025obf} \footnote{Just before the submission of the final version of this paper, the result of the mentioned analysis was published \cite{IceCube:2026plr}, with a final significance of 5.7$\,\sigma$.}. 

In \cite{IceCube:2025pab}, first attempts have been made 
to divide the Galactic Plane into segments in galactic longitude and 
to fit spectral index and flux normalization separately in each segment.
Instead of using a fine-grained spatial template of galactic neutrino emission, the method assumes uniform emission in each segment. It is independent of model assumptions such as those applied in \cite{IceCube:2023ame}.

A quite different approach to search for galactic neutrinos has been chosen in \cite{Kovalev:2022izi}. The authors analyzed public track-like IceCube
events with estimated neutrino energies above 200 TeV. The energy threshold was kept at the same value as in a previous publication searching for a correlation between IceCube neutrino events and blazars. The 200-TeV cut did not have any motivation except data availability. For the Galactic Plane analysis, it was  kept just to avoid any changes, which otherwise had to be motivated and would have added a trial factor. Note, however, that such a cut appears to be counter-intuitive for galactic neutrinos that are expected to have considerably lower energy than those from blazars. The authors examined the distribution of galactic latitude, $b$, of 70 neutrinos. Compared to a diffuse flux, the distribution of the absolute value of |$b$| turned out to be shifted towards lower values, with an apparent excess at |$b$|$ < 20^\circ$.  This  deviates from the hypothesis of isotropy with $4.1\sigma$ significance. The estimated contribution of these events was about one third of the diffuse flux in this energy band.

\begin{figure}[H]

\includegraphics[scale=0.72]{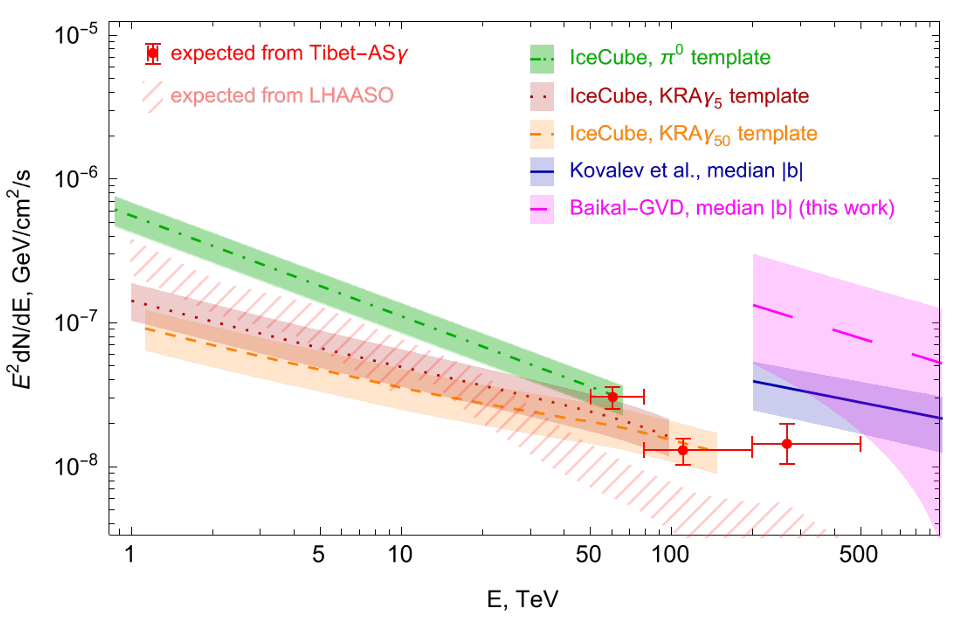}
\caption{
Estimated full-sky spectra of galactic neutrinos
(per one flavor of neutrino plus antineutrino) obtained in
\cite{Baikal-GVD:2024kfx} and in some preceding studies, together with
those expected from observations of diffuse galactic gamma
rays. See the plot legend for notations and details and \cite{Baikal-GVD:2024kfx} for references. Figure taken from \cite{Baikal-GVD:2024kfx}.
}
\label{galactic-200TeV}      
\end{figure}

In a recent publication \cite{Baikal-GVD:2024kfx}, the authors applied the same method to eight cascade events in Baikal-GVD as well as to 12 high-energy contained cascade events (HESE) and 67 track events in IceCube (all with the same energy threshold of 200 TeV). The IceCube tracks are from a more recent compilation of IceCube events (ICECAT) and slightly differ in number and reconstructed energy from the data used in \cite{Kovalev:2022izi}. All the samples show an excess towards small galactic latitudes. The significances for the deviations from isotropy are $2.5\,\sigma$ for the GVD cascades, $2.6\,\sigma$ for the IceCube cascades and $3.1\,\sigma$ for the IceCube tracks. The combined significance is $3.6\,\sigma$. The authors calculated the resulting galactic neutrino fluxes between 200 TeV and 1 PeV according to the three spectral templates mentioned above. The result is shown in Figure~\ref{galactic-200TeV}, together with the IceCube results from \cite{IceCube:2023ame} and 
expectations from Tibet-AS$\gamma$ and LHAASO data assuming a
common origin of neutrinos and photons in proton collisions.
The low-energy extrapolation of the galactic flux obtained in \cite{Baikal-GVD:2024kfx} overshoots both the IceCube data and data derived from gamma rays. Its size is in tension with assumptions of most Galactic Plane models and suggests a wider Milky Way in neutrinos, at least at the highest energy.

Summarizing on the diffuse flux of neutrinos from the Galactic Plane: its discovery at energies $ 100$ TeV can be taken as granted, while the behavior at the highest energy stays an open question. It is expected that high-quality data from KM3NeT with sufficient statistics will appear in a few years and can help in resolving the case, possibly together with track events from Baikal-GVD. KM3NeT also opens the possibility to identify the contribution from individual sources to the total galactic flux.

\section{Search for Point Sources}
\label{point-source-section}

The main goal of large neutrino telescopes is to locate the point-like sources of cosmic neutrinos and obtain insight into their character and the production mechanisms of high-energy cosmic rays. In the following, I will first address searches for steady sources and then those for transient sources.

\subsection{Steady Sources}\label{sec6.1}

Most searches for steady sources try to identify isolated ``point-like'' excesses over the background of atmospheric neutrinos. 


Three classes of events can be used for point sources.

\begin{itemize}
    \item Most searches for point-like sources are based on throughgoing muons, i.e., muons entering the detector from outside and leaving it on the other side. That requires the rejection of all the muons with entrance angles of a few degrees above the horizon since, at smaller zenith angles \footnote{$\theta = 0^{\circ}$ corresponds to vertically downgoing and $\theta = 180^{\circ}$ to vertically upgoing muons.}, the rate of punch-through muons from the atmosphere would dwarf that of neutrino-generated muons. The higher the measured muon energy, the smaller the background of punch-through muons and the larger the angle above the horizon within which neutrino events dominate over background. 
    
    \item Another class of events is ``starting track'' events that are generated by a muon or tau neutrino CC interaction with the vertex contained inside the detector volume. For this class of events the detector is also sensitive to the upper hemisphere. 
    
    \item Last but not least, one can use contained cascade events that, however, suffer from a larger directional uncertainty. 
\end{itemize}

\begin{figure}[H]

\includegraphics[scale=0.60]{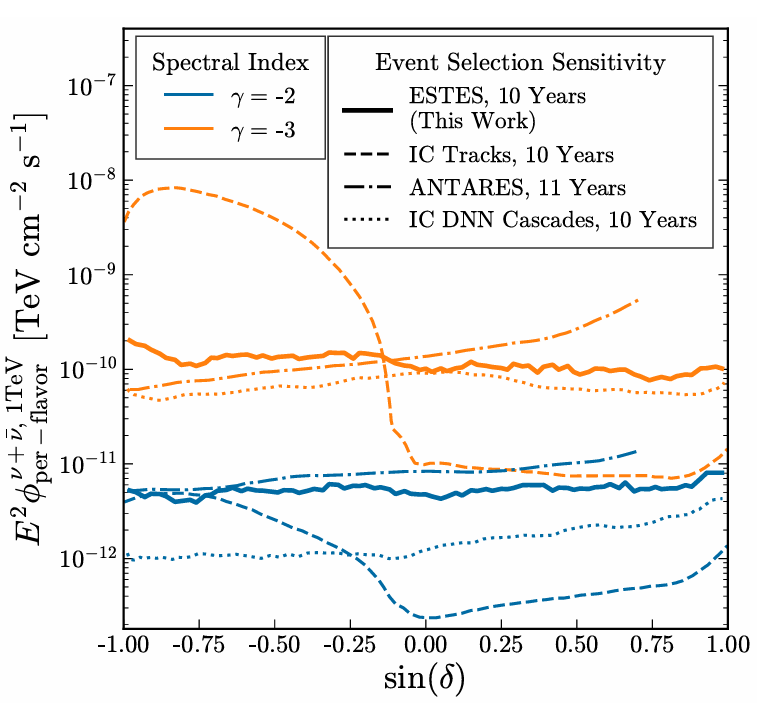}
\caption{
The sensitivity 
 to neutrino and antineutrino per$-$flavor flux at 1 TeV for a steadily emitting point source versus declination for the ESTES analysis (solid lines) and two different source spectral index hypotheses ($\gamma = -2.0$ in blue and $\gamma = -3.0$ in orange). The ESTES sensitivity is compared to that from the 10-year IceCube track sample, the IceCube deep neural network (DNN)$-$selected cascade search and the 11-year ANTARES sample (figure taken from \cite{IceCube:2025zyb}).
}
\label{ESTES-sin-theta}      
\end{figure}

Figure \ref{ESTES-sin-theta} illustrates the situation, with a focus on starting-track events.
The IceCube {\it Enhanced Starting Track Event Selection} (ESTES) is most sensitive to neutrino energies of 1--500 TeV and has a median angular resolution of $1.4^{\circ}$, i.e., worse than for throughgoing tracks. The selection not only rejects atmospheric muons but also suppresses atmospheric neutrinos from the southern sky that are accompanied by punch-through muons from the same air shower (see also the explanation of HESE events in Section \ref{sec4.1}).

The ESTES selection improves IceCube's sensitivity to sources in the southern sky, in particular for galactic sources that are supposed to dominantly produce neutrinos energies below 100 TeV (however, see 
\cite{Baikal-GVD:2024kfx}, where a substantial flux above 200 TeV is claimed). 
The ESTES and cascade sensitivity is approximately uniform over the whole sky, different from IceCube's sensitivity using throughgoing tracks. For a steeper spectrum, the sensitivity abruptly worsens for declinations $\sin\delta < 0^{\circ}$, while, for a flat spectrum, it reaches far into the 
the Southern Hemisphere since, at high energies, the punch-through muon background strongly decreases. The ANTARES sensitivity \cite{Aublin:2019zzn} is more uniform due to the geographic location of the detector.

\begin{figure}[H]

\includegraphics[scale=0.60]{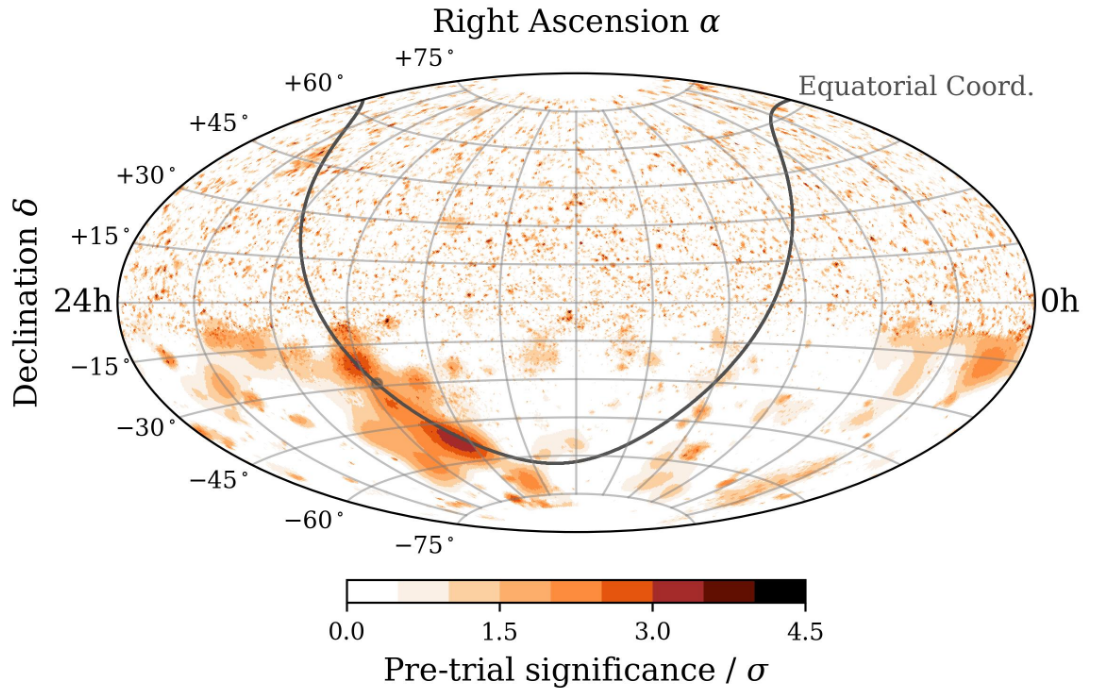}
\caption{Combined tracks and cascades skymap.
All-sky map (equatorial coordinates) of pre-trial significances using
combined cascades and tracks. The solid gray line denotes the Galactic Plane with the dot representing the Galactic Center. The most significant spot in the northern sky aligns with NGC 1068. Figure taken from \cite{IceCube:2025lev}.
}
\label{combined-skymap}      
\end{figure}

Point source searches typically use unbinned maximum-likelihood methods to search for an excess of events at a specified location. The likelihood function is based on the angular distance to the supposed position, the angular
uncertainty and the energy of each event. 
From the energies of events in a given direction, a hypothetical neutrino spectrum
can be estimated. A spectrum harder than that of atmospheric neutrinos enhances the probability for a real source. The result of this method is a map like that shown in Figure~\ref{combined-skymap} \cite{IceCube:2025lev}. 
It is based on a simultaneous fit of different detection channels. The combination of all-sky tracks  (best pointing and sensitivity in the northern sky) with all-sky cascades (superior energy-resolution and sensitivity in the southern sky) results in the most sensitive all-sky source search. The most significant position in the northern sky aligns with the Seyfert 2 galaxy NGC 1068 (see more on neutrinos from Seyfert galaxies below, with \cite{IceCube:2026hzq} as the most recent publication on that issue). 

With a sufficiently large number of simulations one can determine the probability $p$  that 
an excess is just a random fluctuation. 
If the probability of a random match for a given direction (pre-trial) is $p_1$, the post-trial probability
for $N$ independent attempts becomes $p \sim N \cdot p_1$. The ratio
$p/p_1$ is called the penalty factor.
Assuming, for instance, an area of angular uncertainty of about 2 square degrees and including half of the sky
($2\pi$ steradian) in the analysis, the number of quasi-independent directions 
would be $\sim$$10^4$. A pre-trial significance of >$5\,\sigma$ (i.e., $p_1 \sim 10^{-7}$) would
therefore turn to $3\,\sigma$ ($p \sim 10^{-3}$) after correction for trials.

$N$ can be kept small by considering only directions of pre-selected objects. This is called the catalog-search method. Care has to be taken that the catalog list is fixed {\it before} performing the analysis. Performing  several independent list searches with the same data, e.g., selecting in one search radio-loud blazars, in another one X-ray bright sources and in a third one starburst galaxies, leads to difficulties in determining an overall penalty factor. An inherent problem of catalog searches is emphasized in \cite{Troitsky:2021nvu}: Along with the answer to the question, from which sources the excesses of neutrino events are seen, there has to also be an answer regarding why they are {\it not} seen from other sources that are similarly well motivated from an astrophysical point of view. Needless to say, no sources should retrospectively be added if they are only motivated by slight excesses in previous analyses based, e.g., on partial data samples.

\begin{figure}[H]

\includegraphics[scale=0.60]{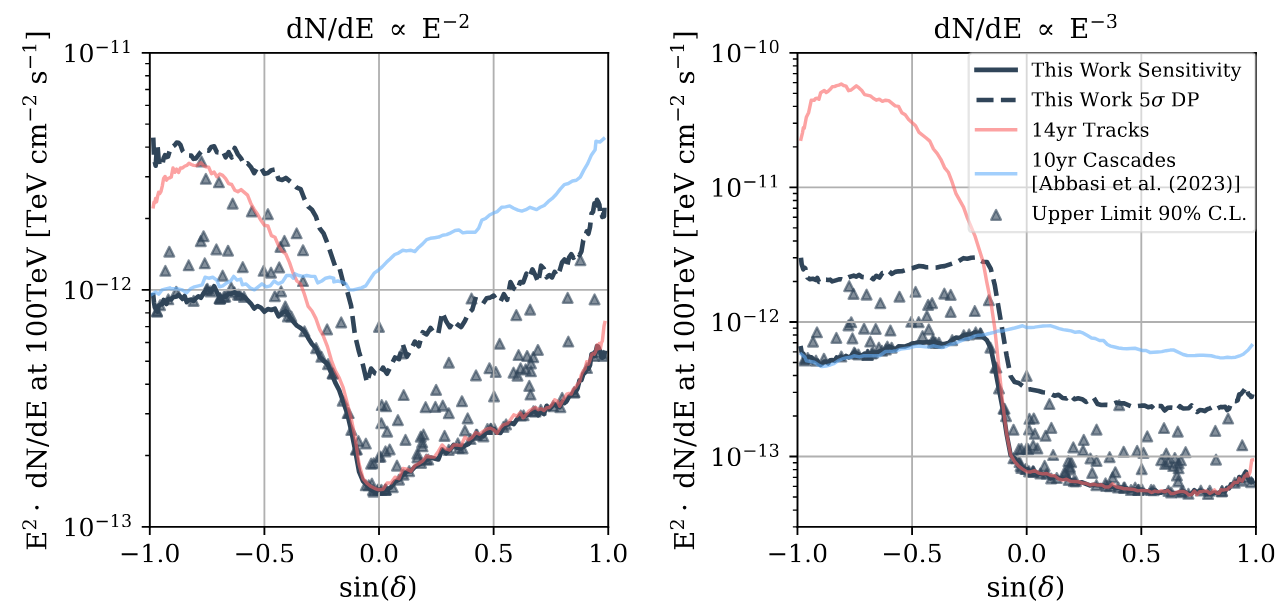}
\caption{
Source list sensitivity, $5\,\sigma$ discovery potential and upper limits as a function of source declination:  The 90\% CL median sensitivity values to sources emitting an $E^{-2}$ spectrum (\textbf{left})
and $E^{-3}$ spectrum (\textbf{right}) for cascades and tracks individually and tracks and cascades combined are shown as full lines. The 90\% CL upper limits for the source catalog sources are
shown as triangles. Note the value for NGC 1068 at declination $\sim$0 (see also Section \ref{catalog}). For sources with the number of ``signal events = 0'', the 90\% CL median sensitivity is plotted instead of the upper limit. The $5\,\sigma$ discovery potential for combined tracks and cascades is shown as dashed line. $dN/dE$ is the per-flavor number of neutrinos ($N$) per neutrino energy ($E$) per area per time.
Figure taken from~\cite{IceCube:2025lev}. 
}
\label{pointsource-2}      
\end{figure}

Figure\,\ref{pointsource-2} shows the sensitivities and 5\,$\sigma$ discovery potentials for a dataset of tracks and cascades and for both components separately for spectral indices $\gamma = -2$ and $\gamma = -3$ \footnote{ {\it Sensitivities} are the fluxes required for 90\% of ``signal plus background''
trials to yield a test statistic (TS) greater than the median TS derived from ``background only'' trials.  The 5\,$\sigma$
{\it discovery potential} is defined as the flux required for 50\% of ``signal plus background'' trials to achieve a TS greater than the $5\,\sigma$ threshold of the ``background only'' TS distribution \cite{IceCube:2025lev}.}.
The two strongest excesses in \cite{IceCube:2025lev} are observed in the directions of the Seyfert galaxy NGC 1068 (Messier 77 in another notation) with a local (i.e., non-trial-corrected) $p$-value of $1.3 \times 10^{-6}$ ($4.7\,\sigma$) and $2.1 \times 10^{-4}$ ($3.5\,\sigma$) when trial-correcting for the 167 sources in the catalog. Its IceCube history dates back to 2022 (see Section \ref{catalog}). The second strongest excess came from the direction of the blazar PKS 1424+240, with a local $p$-value of $10^{-3.5}$. 
The blazar TXS 0506+056 is on the fifth place in this analysis, with a local $p$-value of $10^{-2.6}$. It was first noticed in its capacity as a {\it transient} source; see Section~\ref{transient-subsection} below.

\begin{figure}[H]

\includegraphics[scale=0.72]{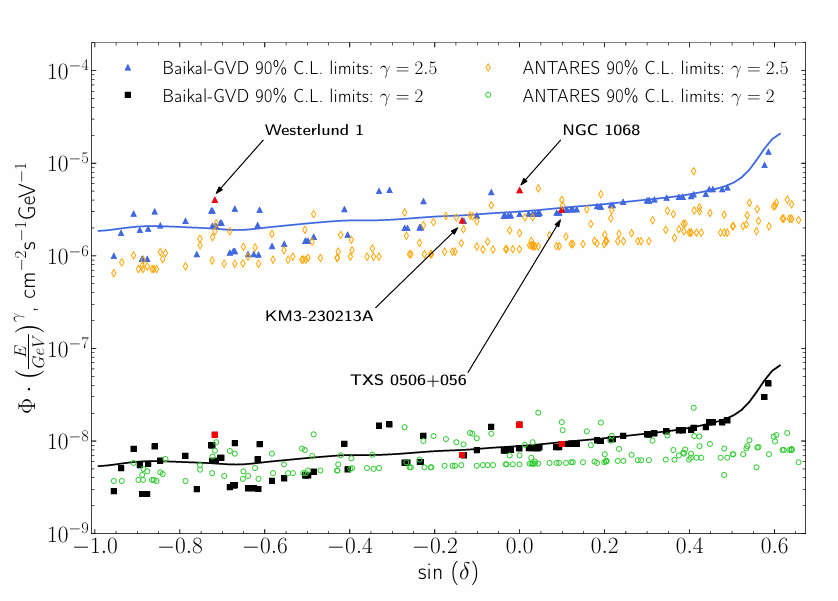}
\caption{
Point 
 source sensitivity (median expected 90\% CL upper limit, solid color lines) and 90\% CL flux upper limits obtained for a catalogue of 92 source candidates using the Baikal-GVD
track-like event sample collected between April 2019 and March 2024 (filled squares and triangles) for two assumed spectral indices $\gamma$. The ANTARES data \cite{ANTARES:2025wft}  (empty circles) are shown for comparison. The symbols for four particular interesting sources are drawn in red. Figure taken from \cite{GVD:2026cep}. 
}
\label{B-and-A}      
\end{figure}

Quite recently, the Baikal-GVD collaboration published their first results from a point source search with track-like events; see Figure \ref{B-and-A}, taken from \cite{GVD:2026cep}.

Since tracks in the GVD analysis have been reconstructed within single clusters only and a cluster is only 120\,m in diameter, the angular error significantly worsens towards the horizon. Since the average number of fired OMs is smaller than for vertical trac\mbox{ks, th}e energy threshold rises. Data have been taken with the partially completed detector between
April 2019 and March 2024. 
The authors have searched for muon neutrino fluxes from 92 astrophysical objects of interest. Different from ANTARES and IceCube, they used a $\chi^2$-based track reconstruction method and
 a cut-based analysis. The obtained limits are of the same order as those set by ANTARES (based on approximately the same number of events).  The angular resolution for nearly vertical tracks is $\sim$0.5$^{\circ}$.   With the inclusion of multi-cluster tracks the GVD sensitivity over the full declination range is expected to improve substantially.
\unskip



\subsubsection{Catalog Searches}
\label{catalog}

There are a multitude of IceCube searches based on catalogs, e.g., for galaxy clusters~\cite{IceCube:2021abh}, cores of active galactic nuclei (AGN) \cite{IceCube:2021pgw}, 1FLE blazars \cite{Abbasi:2022uox},
magnetars \cite{IceCube:2023ugc}, hard X-ray AGN \cite{IceCube:2024ayt}, 
or X-ray bright Seyfert galaxies \cite{IceCube:2024dou} --- just to mention publications of the last five years. Most of them gave upper limits only.  

\begin{figure}
\includegraphics[scale=0.40]{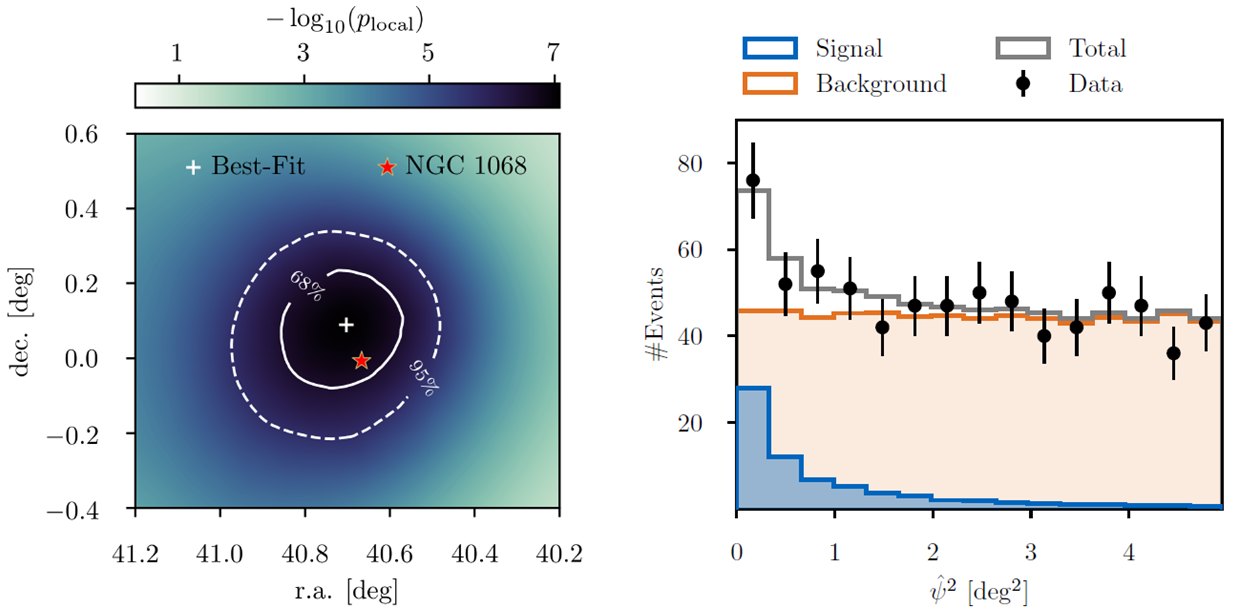}
\caption{
The sky region around the most significant spot in the Northern Hemisphere and NGC 1068. The left plot shows a fine scan of the region around the hottest spot. The spot
itself is marked by a yellow cross, and the red star shows the position of NGC 1068. In addition,
the solid and dashed contours show the 68\% (solid) and 95\% (dashed) confidence regions of
the hotspot localization. The right plot shows the distribution of the squared angular distance
between NGC 1068 and the reconstructed event direction. From Monte Carlo simulations one estimates the
background (orange) and the signal (blue) assuming the best-fit spectrum at the position of
NGC 1068. The superposition of both components is shown in gray and provides an excellent
match to the data (black). 
Figure taken from~\cite{IceCube:2022der}. 
}
\label{NGC-excess}      
\end{figure}

The only compelling observation \footnote{Note that “compelling” or “compelling evidence” is not well defined. It was used by the IceCube collaboration for NGC 1068 to emphasize the stronger trust in the signal, which appeared to become more significant with more statistics and with improved analysis methods.} of a steady single source concerns the above-mentioned NGC 1068, a Seyfert galaxy of Type 2. This is the source with declination $\delta \sim 0$ standing out in the data shown in Figure\,\ref{pointsource-2}. It was first observed in \cite{IceCube:2022der} with a significance of 
$4.2\,\sigma$; see Figure \ref{NGC-excess}. The $4.2\,\sigma$ value is the result of a catalog search based on 110 sources. The significance derived from a full-sky scan was $2.0\,\sigma$. 

 The mentioned catalog had been used in previous analyses and expanded (including, among other sources, NGC 1068), just before the analysis published in \cite{IceCube:2022der} was performed.  So, as noted in \cite{Troitsky:2021nvu}, the term ``a priori fixed'' in the context of NGC 1068 may not be fully justified. On the other hand, its excess from previous analyses was not noticeable, so the {\it a posteriori} inclusion can presumably not be counted as bad practice. The spectral index derived for the NGC 1068 excess events came out to be $3.2 \pm 0.2$, which is not much harder than that of atmospheric neutrinos ($\sim$3.7). 

\begin{figure}[H]

\includegraphics[scale=0.7]{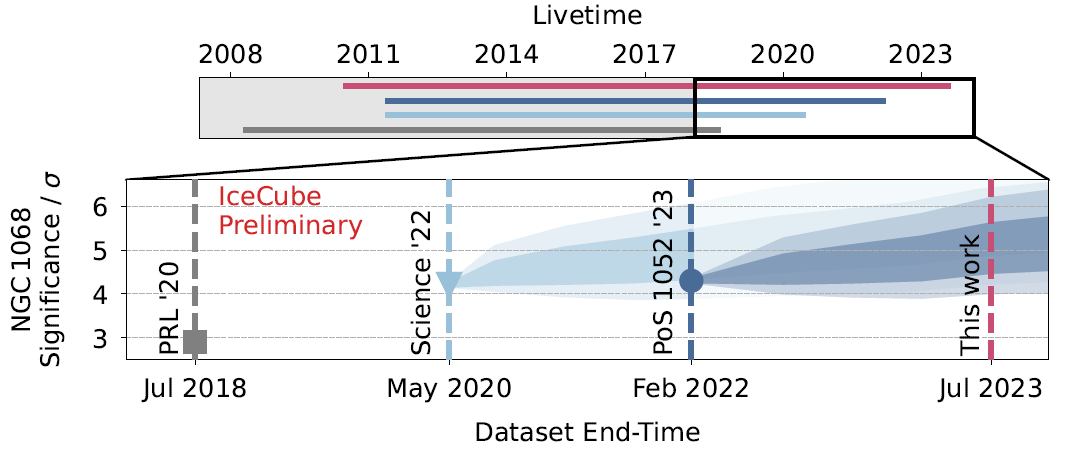}
\caption{
The upper panel compares the livetimes of the various datasets used to measure the neutrino
flux from NGC 1068. The lower panel zooms in on the end time of the datasets and shows the increasing
significance of the emission over the years. The result from July 2018 (gray square) used different data
processing and analysis methods. Starting from the May 2020 publication (light blue triangle), the rise in significance is
due to the increased statistics.  The significance from 2022 on (dark blue dot) is compatible with the expected statistical fluctuations within $1\,\sigma$ (dark shaded bands). The lighter shaded bands indicate the 95\%
containment of the fluctuations. The actual value found in this work (July 2023, not shown in the figure) is $4.3\,\sigma$.
Figure taken from \cite{IceCube:2023wid}. 
}
\label{NGC-over-years}      
\end{figure}

Figure \ref{NGC-over-years} shows the development of the global significance of the NGC 1068 signal over the years \cite{IceCube:2023wid}. NGC 1068 was included in the source list after the 2018 analysis, but that analysis did not yet show a particularly noticeable excess. In the most recent publication~\cite{IceCube:2026hzq}, NGC 1068 was observed with a global significance of $4.0\,\sigma$ --- a bit on the lower side of what could have been expected from statistical fluctuations only. However, the significance may change due to factors beyond statistics that are not always obvious to the reader: new reconstruction procedures or improved ice parameters used in the reconstruction that lead to slightly different arrival directions of a few tenths of a degree for tracks and several degrees for cascades; inclusion of different event topologies; inclusion or exclusion of single events due to changed selection criteria, etc. So, while in 2020 \cite{IceCube:2022der} one might have been reluctant to accept NGC 1068  as a compelling source and have had doubts about how meaningful detailed theoretical models were in their function to explain the data, the data accumulated since then certainly underpin the quality of the excess as ``compelling evidence''.

\begin{figure}[H]

\includegraphics[scale=0.98]{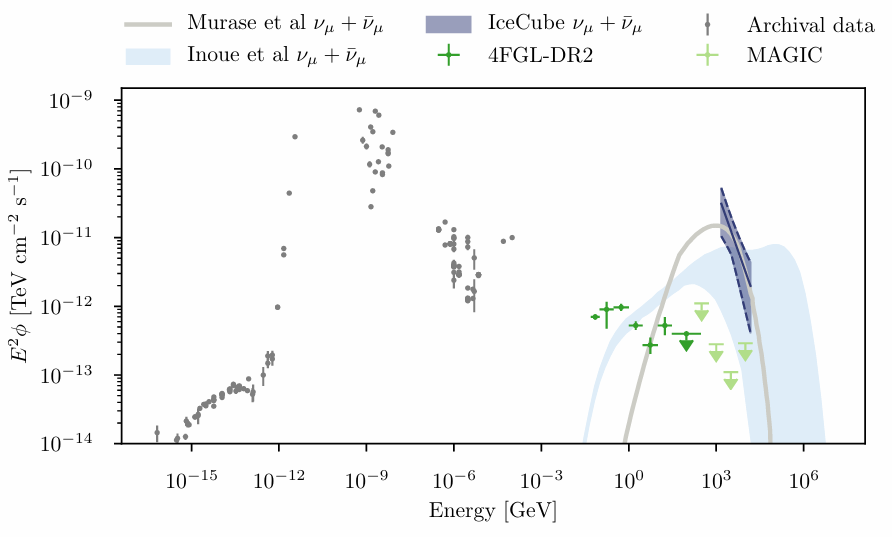}
\caption{
{Spectral energy distribution of NGC 1068.}
Gray 
 points show publicly available
multi-frequency measurements. Dark and light green error bars refer to gamma-ray
measurements from Fermi-LAT and MAGIC, respectively. The solid dark blue line
shows the best-fit neutrino spectrum, and the corresponding blue band covers all power-law neutrino fluxes that are consistent with the data at 95\%\,CL. It is shown in the energy range
between 1.5 TeV and 15 TeV where the flux measurement is well constrained. Predictions for two AGN core models for neutrino emission  \cite{Inoue:2019yfs,Murase:2019vdl} are shown for comparison: the light blue shaded region and the gray line.
Figure taken from \cite{IceCube:2023wid}. 
}
\label{NGC-MAGIC}      
\end{figure}

NGC 1068 is one of the closest and best-studied Seyfert 2 galaxies and has been discussed as
a potential source of high-energy neutrinos for a while. Notably, no gamma rays of $>100$ 
 GeV are observed from the direction of NGC 1068
(see Figure \ref{NGC-MAGIC}). 
They should be produced together with neutrinos but can lose energy 
in an optically thick environment. X-ray photons generated through photon Comptonization in the hot plasma above the disk absorb gamma rays but also provide a target for the production of neutrinos. Figure\,\ref{NGC-MAGIC} compares the neutrino flux measurement with estimations of two such models. 

\begin{figure}[H]

\includegraphics[scale=0.65]{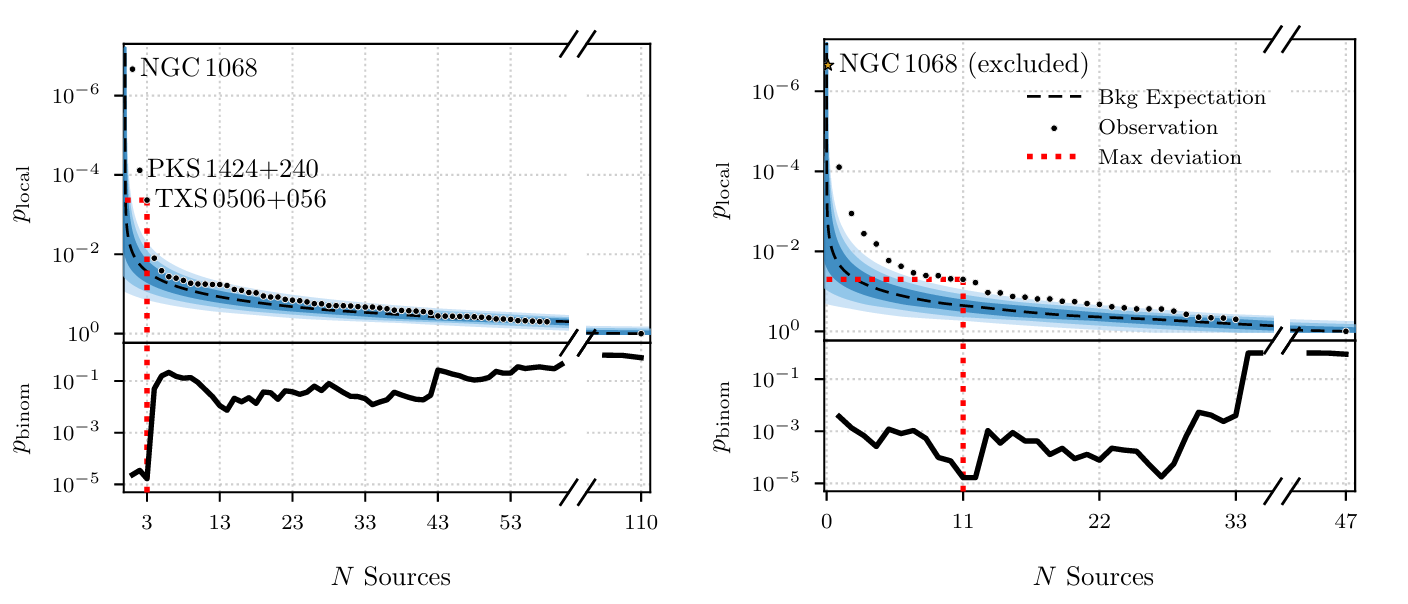}
\caption{
Visualization of the binomial test excess for the list of 110 gamma-ray emitters (\textbf{left}) and 47~X-ray-bright AGN (\textbf{right}). In the upper panels, the local $p$-values for each source are shown, ordered from the lowest to the highest (black points), the background expectation (dashed black line), and its $1\,\sigma$, $2\,\sigma$ and $3\,\sigma$ Poissonian uncertainties (shaded blue bands). The lower panels show the binomial probability for each subset of sources. The most significant excess is highlighted by dotted red lines. To preserve readability, the plot is truncated at the point where sources reach a $p$-value of 1.0, beyond which the
binomial probability is also equal to 1.0.
Figure taken from \cite{IceCube:2026hzq}. 
}
\label{binomial_test}      
\end{figure}

In the mentioned most recent analysis \cite{IceCube:2026hzq} --- based on IceCube track events recorded over 13.1 years --- NGC 1068 remains the most significant neutrino source among 110 preselected sources and at the same time is spatially compatible with the most significant spot in the northern sky. Its energy spectrum,
as measured in \cite{IceCube:2026hzq},  is described by an unbroken power law with spectral index $\gamma = 3.4\pm0.2$. As mentioned above, next to NGC 1068 follow
two blazars, PKS 1424+240 and TXS 0506+056. The dominance of these three among 110 sources becomes obvious when performing a binomial test. Such tests can be used to identify groups of sources that may show weak individual signals but yield a significant collective excess. The local $p$-values of the $N$ sources are sorted in ascending order. For each rank $k = 1,...,N$ one calculates the probability of observing $k$ or more sources with $p$-values below
the $k$-th smallest value in the list in a background-only scenario and compares that to the actual value. The outcome of the test consists of the number of sources $k$ that provide the most significant collective excess above the background, i.e., the smallest binomial $p$-value among all the tested cases. Figure \ref{binomial_test}, left, shows that for the 110 sources considered.

Motivated by the disparity between gamma-ray and neutrino emissions and the high X-ray luminosity of NGC 1068, the authors of \cite{IceCube:2026hzq} selected 47 X-ray bright Seyfert galaxies from the Swift/BAT spectroscopic survey for an alternative catalog search. None of these sources had been included in the list of 110 gamma-ray emitters.  Figure \ref{binomial_test}, right, shows the results of the binomial test for these 47 sources. A $3.3\,\sigma$ excess is observed
for an ensemble of 11 sources (with NGC 1068 excluded from the sample!). 
Among the contributing 11 AGN are both Seyfert 1 (e.g., NGC 4151 and NGC 7469) and Seyfert 2 (e.g., NGC 1068 and CGCG420-015) galaxies, suggesting that the level of obscuration by the corona (stronger for Seyfert 2) does not significantly impact the likelihood of neutrino emission. 

\vspace{8mm}

\unskip

\begin{figure}[H]

\includegraphics[scale=0.8]{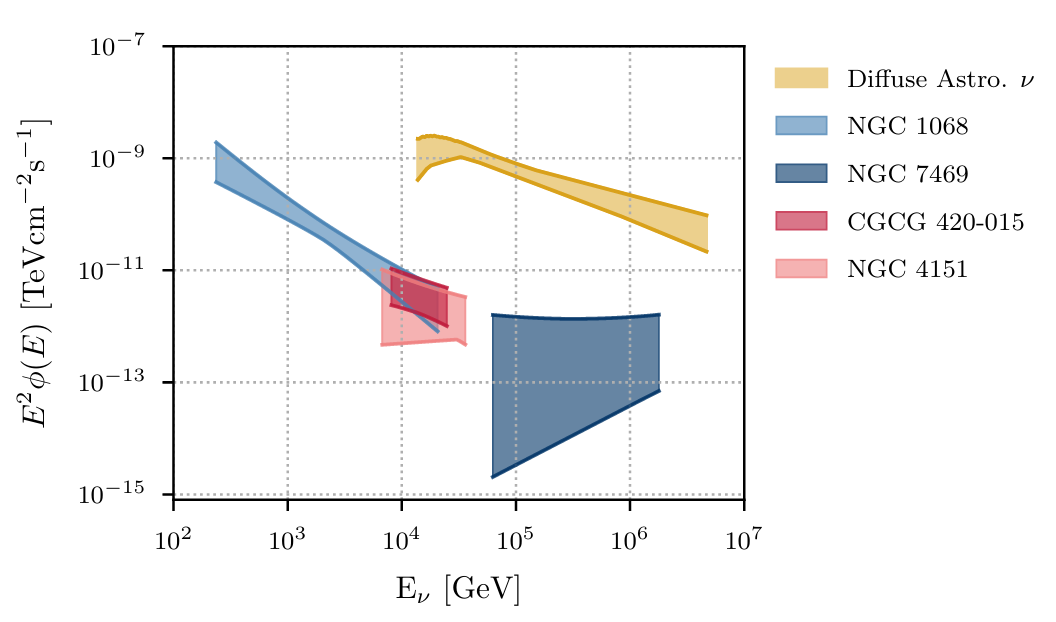}
\caption{
Best-fit neutrino all-flavor fluxes with their $1\,\sigma$ uncertainty and constrained in the 95\% CL energy range of the top four
sources in the list of X-ray bright AGN. The yellow band represents the most recent measurement of the diffuse astrophysical
neutrino flux. The energy range is constrained at the 90\% CL, while the flux normalization is shown with a 68\% CL. Figure taken from \cite{IceCube:2026hzq}. 
}
\label{4-strongest}      
\end{figure}

Figure \ref{4-strongest} shows the fluxes for the four strongest X-ray bright AGN (here {\it including} NGC 1068) and compares them to the diffuse flux measured by IceCube.

\vspace{1cm}

\newpage

\subsubsection{Blazars}
\label{blazars-subsection}

Blazars, with their jets pointing in the direction of Earth, have been considered to be the most promising candidates for neutrino detection for a long time. Directionality and strong Doppler boosting (and less $\gamma$ obscuration, as expected for Seyfert  galaxies) make them an ideal candidate for being observed in neutrinos. From Fermi-LAT observations one knows that they are the most abundant sources of gamma rays, constituting about 80\% of the entire extragalactic population. They are also expected to emit neutrinos. 

\begin{figure}[H]

\includegraphics[scale=0.84]{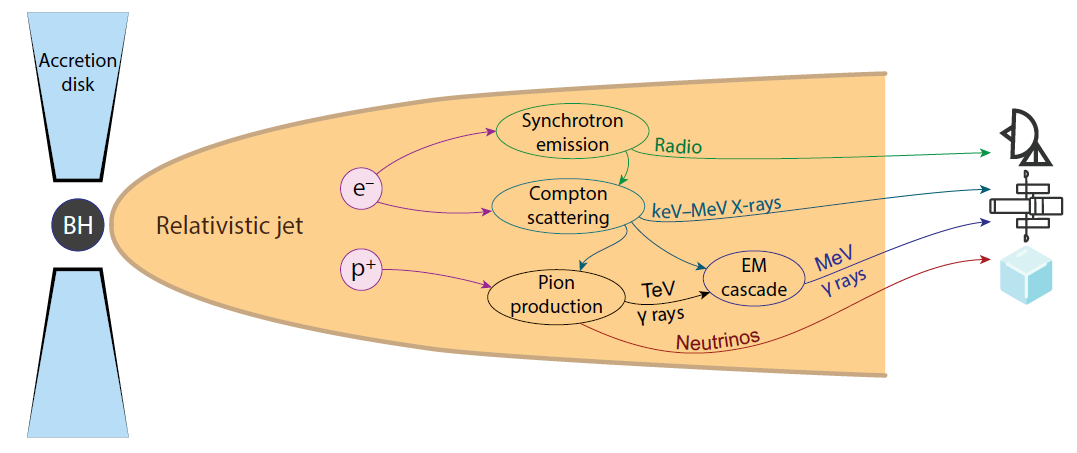}
\caption{
Production of neutrinos and relevant electromagnetic (EM) emission in relativistic jets of blazars at parsec scales. Figure taken from \cite{Plavin:2020mkf}. 
\copyright{AAS}. Reproduced with permission.
}
\label{Blazar-basics}      
\end{figure}

The spectral energy distribution (SED) of their electromagnetic radiation shows two broad distinctive peaks:
a low-energy peak between radio and X-ray energies, which is due to synchrotron emission of energetic electrons,
and a high-energy peak at gamma-ray energies, which can
be explained by inverse Compton scattering and hadronic interactions. Figure~\ref{Blazar-basics} illustrates the relevant processes leading to electromagnetic and neutrino~production.

\vspace{1.3cm}

\vspace{1.3cm}

On a source-by-source basis, the $\nu/\gamma$ correlation may vary considerably. It depends, e.g., on the energy of the target photons, and of course on the level of gamma-obscuration by surrounding matter. One strategy to cope with this uncertainty is just to {\it stack} the signals from the directions of a large sample of blazars based on the idea that at least their average properties are related to neutrino production. 

\begin{figure}[H]

\includegraphics[scale=0.85]{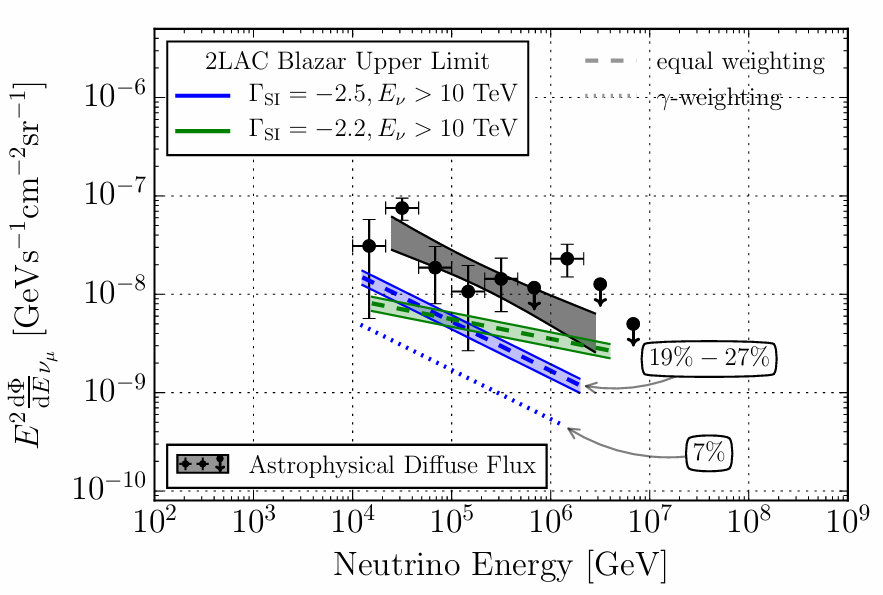}
\caption{The 90\% CL flux upper limits for all 2LAC blazars in
comparison to the observed astrophysical diffuse neutrino flux. The
2015 combined diffuse neutrino flux results are
plotted as the best-fit power law with spectral index
$-$2.5 and a differential flux unfolding using 68\% central and
90\% UL confidence intervals. The flux upper limit is shown using
two weighting schemes (see text) for a power law with spectral index $-$2.5
(blue). Percentages denote the fraction of the upper limit compared
to the astrophysical best-fit value. The equal-weighting upper limit
for a flux with a harder spectral index of $-$2.2 is shown in green.
Figure taken from \cite{IceCube:2016qvd}. \copyright{AAS}. Reproduced with permission. }
\label{blazar-IC}      
\end{figure}

A stacking approach was chosen in \cite{IceCube:2016qvd}. The first three years of IceCube data taking (2009--2012) were used to search for neutrino emission from 862 blazars included in the second Fermi-LAT AGN catalogue (2LAC). The sources were {\it (a)}  
 not weighted and {\it (b)} weighted according to their gamma-ray activity. Excesses of $1.6\,\sigma$ and $0.4\,\sigma$ were observed for the unweighted and weighted sum of all blazars.
In Figure\,\ref{blazar-IC}, the corresponding flux limits from all 862 blazars are
compared to the observed astrophysical diffuse neutrino flux. Concluding from this analysis, the maximum contribution of blazars to the diffuse flux of cosmic neutrinos is 27\%.
Note that, for this analysis, a spectral index of $-$2.5 was assumed. Changing the spectral index of the tested flux to $-$2.2 weakens the constraint by about a factor of two, i.e., from 27\% to 50--60\%.

\vspace{1cm}

\begin{figure}[H]

\includegraphics[scale=0.6]{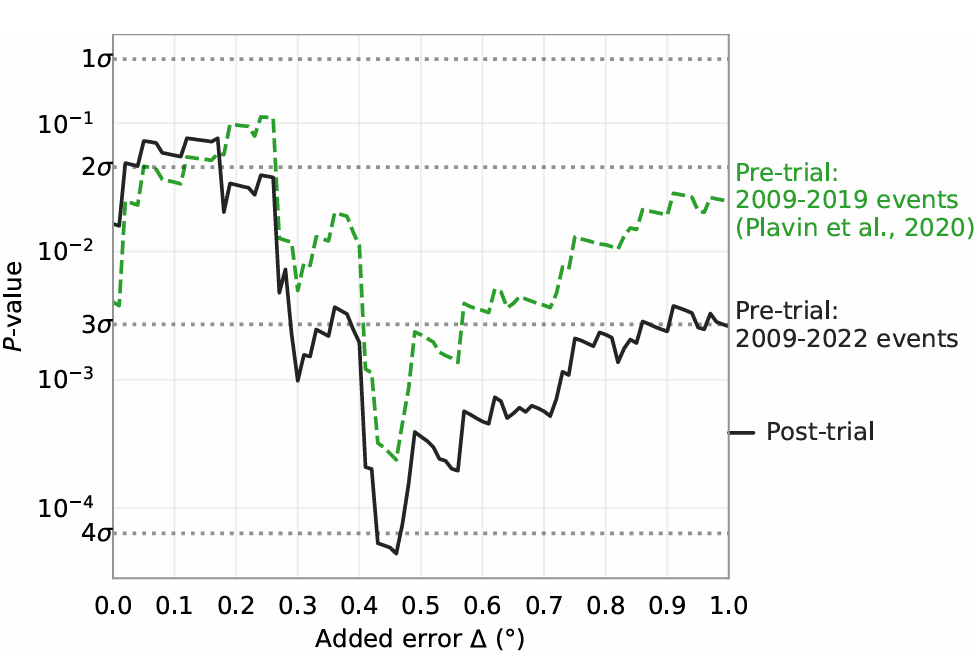}
\caption{The $p$-values for bright VLBI blazars being spatially correlated with IceCube high-energy neutrinos. The x-axis gives the added error $\Delta$.
The two curves correspond to the sample of 56~neutrino events analyzed in \cite{Plavin:2020emb} and the sample of 71 events from \cite{Plavin:2022oyy}, respectively.
The post-trial $p$-values for these events are $p = 2\cdot10^{-3}$ and 
$p = 4\cdot10^{-4}$ (indicated by the post-trial mark in the figure), respectively. 
Figure taken from \cite{Plavin:2022oyy}. 
}
\label{VLBl}      
\end{figure}

An unconventional approach was chosen in \cite{Plavin:2020mkf}.
The authors used radio data from very long baseline interferometry (VLBI) to define a sample of 3412 AGN selected according to their radio emission. Parsec-scale emission from blazars indicates energetic processes in the jet and Doppler boosting effects. Fifty-six IceCube tracks with energies $\geq$\,200 TeV were used, with the threshold chosen beforehand and not optimized. The events were registered between 2009 and 2019. In their analysis, the authors used the angular error ellipses given by the IceCube collaboration but added a free parameter $\Delta$ to those errors. Looking for a correlation  between directions of the AGN positions and those of the IceCube events they found $p$-values compatible with no correlation at all. Varying $\Delta$, however, the pre-trial $p$-value decreased and reached a minimum $\sim$$3\cdot10^{-4}$ at $\Delta \sim 0.45^{\circ}$. In \cite{Plavin:2022oyy} they repeated the procedure after adding 14 more IceCube events registered between 2019 and 2022. The effect became even stronger (see Figure\,\ref{VLBl}) and reached a $p$-value slightly less than $4\cdot10^{-4}$ ($3.6\,\sigma$). Performing the same analysis on lower-energy events and combining the results of the high-energy and lower-energy analyses, they obtained $p = 1.9\cdot10^{-5}$ ($4.3\,\sigma$). If confirmed, this effect would not only support the mechanisms for neutrino generation developed in \cite{Plavin:2022oyy} but also require an underestimation of IceCube's track angular errors by about half a degree (or about a factor 2 in coverage area).  

Cross checks within the IceCube collaboration using slightly different event and catalog samples \cite{IceCube:2023htm} led to a much smaller effect than claimed in \cite{Plavin:2020emb}.  Efforts  are underway to further investigate the effect of error scaling using more data and possibly also modified catalogs. Whatever the outcome of these studies, the present situation nicely demonstrates how deeply the significance of an observation depends on alleged ``subtleties'' like catalog definitions, event selection and reconstruction errors! Everything below 4\,$\sigma$ can be taken as an ``indication'' at best.

As of now, no significant correlation between radio-loud AGN and neutrino emission was found in analyses of the IceCube collaboration. In \cite{IceCube:2021imv}, for instance, a search was performed for a correlation between 275 IceCube {\it alert events} (see below for details of the alert program) and 2089 blazars from the Fermi-LAT 4LAC-DR2 catalog as well as 3413~AGNs from the {\it Radio Fundamental Catalogue} --- without any statistically significant effect. 

In \cite{Buson:2022fyf,Buson:2023irp} a different catalogue than in \cite{Plavin:2020mkf,Plavin:2022oyy} was used, the 5th data release of the Roma-BZCat
catalogue (5BZCat). The threshold for the neutrino events is set to 100~TeV instead of 200 TeV. Similar to \cite{Plavin:2022oyy}, the authors calculate the degree of blazar--neutrino correlation as a function of {\it a posteriori} parameters, in this case not one but two of them. The first is the local probability of a neutrino spot to be astrophysical, $L_{min}$ and the second a parameter they call the association radius $r_{assoc}$, which is the minimum distance between two clusters to be counted as separate. In \cite{Buson:2022fyf} only the Southern Hemisphere is investigated. A minimum chance probability of
$p = 3 \cdot 10^{-7}$ is achieved with the optimum set of $L_{min}$ and $r_{assoc}$,
yielding a post-trial chance probability $p = 6 \cdot 10^{-7}$ (based on 1177 blazars,
19 hotspots found in the neutrino sample, with 10 of them matching a 5BZCat position).
In \cite{Buson:2023irp}, an analogue analysis was repeated for the northern sky.
A correlation between blazars and northern neutrino data at the pre-trial probability of $p = 5.1 \cdot 10^{-4}$ with a
post-trial probability of $p = 6.8 \cdot 10^{-3}$ was found. Combining the post-trial probabilities observed for the southern and northern experiments yields a global post-trial chance probability of $ p = 3.6 \cdot 10^{-9}$.
The authors conclude that this was at an ``unprecedented level of confidence, providing the observational evidence that blazars are astrophysical neutrino factories and hence, extragalactic cosmic-ray accelerators''.


\subsection{Transient Sources}
\label{transient-subsection}

In 2016, IceCube started its alert program: a real-time system to detect events with a high probability to be of extraterrestrial origin and disseminate information about those events worldwide with a latency of seconds to minutes (see Section \ref{Synergies-section} for details). These alerts enable the global astronomy community to instantly observe the same region of the sky and --- in the case of finding an active source --- support the hypothesis that the event is extraterrestrial and in addition emerges from a transient source flaring in electromagnetic radiation or gravitational waves. Initially limited to track events, since 2020, the alert program also includes cascade events. Two types of alerts are being released: gold alerts (about 10 per year, with more than 50\% likelihood to be of cosmic origin) and bronze alerts (about 20 per year, with 30\%--50\% likelihood  to be of cosmic 
origin). The likelihood is mainly determined by the energy.

In 2017, the alert system led to the first identification of a blazar as a neutrino source~\cite{IceCube:2018dnn}.
On 22 September 2017, a neutrino with an energy of approximately 290~TeV triggered the alert system (IceCube-170922A). Its arrival direction was consistent with the location of a known gamma-ray blazar, TXS 0506+056. The alert was followed by an extensive multi-wavelength campaign, ranging from radio frequencies to gamma rays. The campaign revealed that TXS 0506+056 was in a flaring state.  
Based on the directional coincidence of IceCube-170922A with TXS 0605+056 and the information about the flaring state of that object, a {\it chance} correlation between that high-energy neutrino and the direction of TXS 0506+056 could be rejected at the $2\,\sigma$ level.

\begin{figure}[H]
\includegraphics[scale=0.7]{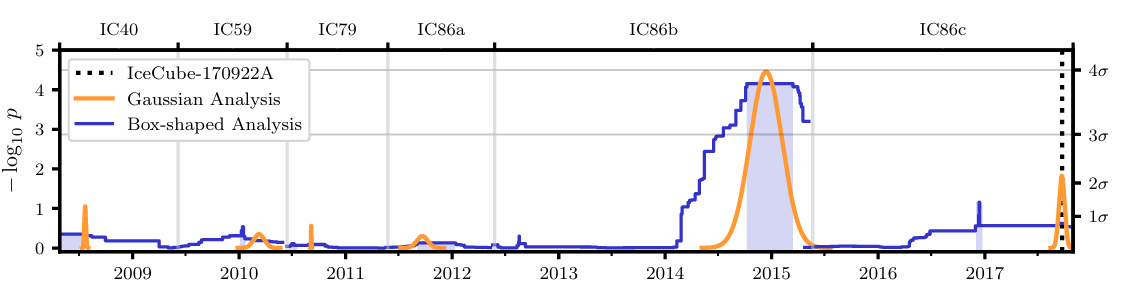}
\caption{ 
{Time-dependent 
 analysis results.} The orange curve corresponds to the analysis
using the Gaussian-shaped time profile. The central time $T_0$ and width $T_W$ are plotted for the
most significant excess found in each period, with the $p$\,value of that result indicated by the
height of the peak. The blue curve corresponds to the analysis using the box-shaped time profile.
The curve traces the outer edge of the superposition of the best-fitting time windows (durations
$T_W$) over all times $T_0$, with the height indicating the significance of that window. In each period, the most significant time window forms a plateau, shaded in blue. The large blue band centered near 2015 represents the best-fitting 158-day time window found using the box-shaped time
profile. The vertical dotted line on the right indicates the time of the IceCube-170922A event. Figure taken from \cite{IceCube:2018cha}.
}
\label{TXS}      
\end{figure}

Prompted by that result, the IceCube collaboration scanned 9.5 years of data 
for transient emission at the position of TXS 0506+056 \cite{IceCube:2018cha}.
And indeed, for a period between September 2014 and March 2015, an excess of high-energy neutrino events over the atmospheric background at that position was found. 
The time-dependent analysis, using the same basic formulation of the
likelihood, searched for clustering in space {\it and time} by introducing an additional time profile. Two generic profiles were tested: a Gaussian-shaped time window and a box-shaped time window; see Figure\,\ref{TXS}.  Allowing for time-variable flux, that analysis constituted $3.5\,\sigma$ evidence for neutrino emission from the direction of TXS 0506+056 prior to the 2017 flaring episode. 

Archival data from Fermi-LAT revealed that TXS 0506+056 was in a low-gamma-ray state during the neutrino excess in 2014-15. No good agreement between the observed neutrino signal and other wavelength observations was obtained, including later gamma outbursts, which were all {\it not} accompanied by any neutrino signal, which is a challenge for model builders. The best descriptions have been obtained for hadronic scenarios where either gamma rays are absorbed within the source or neutrinos and gamma rays stem from different zones in the jet. Two-zone models often include a combination of leptonic and hadronic processes. The inner zone would produce high-energy gamma rays and neutrinos via proton-synchrotron or photopion processes, while the outer zone produces lower-energy radiation (X-rays/optical) via synchrotron or synchrotron self-Compton (SSC) mechanisms (see  Figure\,\ref{Blazar-basics} for the different mechanisms). An alternative attempt to describe TXS 0506+056 with a single-zone model has been presented, for instance, in \cite{Keivani:2018rnh}.

\begin{figure}[H]

\includegraphics[scale=0.8]{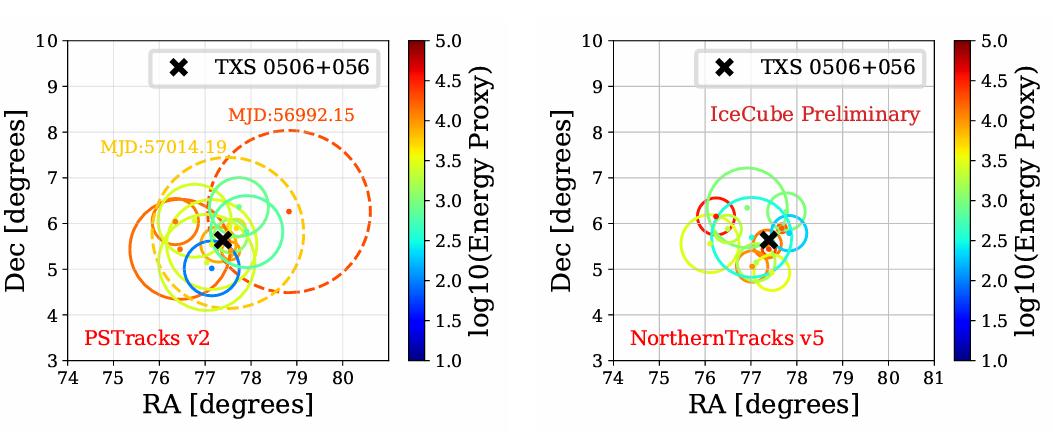}
\caption{Locations 
 of the events that contribute most to the 2014 flare in  the 2017 analysis (\textbf{left}) and in the 2023 analysis (\textbf{right}). The 2 cascade events have been omitted in the 2023 analysis.
The circles represent the $1\,\sigma$ angular error regions for each event, and the color corresponds to the event energy proxy.
 Figure taken from \cite{IceCube:2023oua}.
}
\label{TXS-2017-2023}      
\end{figure}

The identification of TXS 0506+056 is not on such solid ground as that of NGC 1068. While the latter was declared as ``compelling evidence'', the former was only given an “evidence” rating (without ``compelling''). Not only was the significance lower than for NGC but, in a later analysis, the TXS significance for the 2014/15 effect also changed to lower values than in \cite{IceCube:2018cha}. Compared to the initial analysis, in a later analysis \cite{IceCube:2023oua}, numerous updates to event processing and reconstruction, as well as improvements to the statistical methods, were realized. Two cascade events were omitted in the new analysis due to their significantly worse angular resolution. Figure\,\ref{TXS-2017-2023} demonstrates the different outcomes in the first analysis and the latest analysis. The most probable angles and the energy proxies have slightly changed. The angular errors for most events have significantly decreased. Note, however, that this does not necessarily mean that the new locations are ``better'', at least as long as it is not fully clear how well systematic effects are under control. See  \cite{Troitsky:2021nvu} for a more extensive discussion of this and similar effects.
The new analysis yields a $p$-value of $10^{-3}$ compared to $7\,\cdot\,10^{-5}$ in the first analysis and a flux normalization factor of 2 smaller.

\vspace{5mm}

The existence of a real neutrino flare in 2014/15 can never be ``disproved'' since it was a transient event. What counted most, however, in spite of the moderate significance, was its great encouragement to the community of neutrino astronomers. Also, for theoreticians, TXS 0506+056 became a valuable test object to develop more sophisticated models, notably two-zone models, which could explain the multimessenger panorama of this and other objects. Needless to say, the TXS case also had a positive political impact, eventually to more projects than IceCube alone.

Another approach to identify the transient behavior of blazars was chosen in~\cite{Plavin:2020emb}.
The authors correlate data taken with the radio telescope RATAN-600 to IceCube neutrino events. RATAN-600 with its wide field of view is located in the Russian part of the Caucasus and has been monitoring a sample of AGN at 1--22 GHz since late 1980. The AGN have been selected according to their VLBI flux density. Observations at the highest RATAN frequency of 22 GHz have been used
since  flares are typically more pronounced at shorter
wavelengths. The IceCube neutrino data used were taken in 2009--2019. 
For 18 AGN close to neutrino events, the authors calculate the ratio of RATAN-600 flux densities averaged over a 0.9 year window to the average flux density outside that window. Plotting this ratio versus the time lag $t$ to a neutrino event, they observe a pronounced peak at $\sim$$\pm 0.5$ years around $t$\,=\,0. 
The post-trial $p$-value of 5\% is not very large but at least suggestive, indicating that periods of increased radio emission at high frequencies may correlate with enhanced neutrino emission.

\vspace{7mm}

\subsubsection{Tidal Disruption Events (TDEs)}
\label{TDE-subsubsection}

On 1 October, 2019, the IceCube reported the detection of a 
$\sim$200 TeV neutrino (IC191001A) of the ``gold alert'' type, with an estimated 59\% probability to be of cosmic origin \cite{Stein:2020xhk}. Seven hours later, the Zwicky Transient Facility (ZTF) targeted the corresponding sky region. 
ZTF reported a likely connection of that neutrino to a tidal disruption event with strong radio emission (AT2019dsg, where AT stands for Astronomer's Telegram).

Tidal disruption events (TDEs) occur when stars pass close to supermassive black holes and are torn apart by gravitational tidal forces. The stellar debris forms an accretion disk around the black hole, resulting in a bright electromagnetic flare that can last for months or years.  TDEs are supposed to be sources of ultra-high-energy cosmic rays, gamma rays and neutrinos, notably those with relativistic jets. 
With typically less than two radio-emitting TDEs in the entire northern sky at any one time, the probability of finding a time correlation between a radio-detected TDE and such an energetic neutrino by chance was calculated to be smaller than 0.5\%. 
AT2019dsg turned out to have the second highest bolometric energy flux of all seventeen TDEs detected by ZTF until October 2019. From the TDE brightness distribution, the probability of finding a TDE at least as bright as AT2019dsg by chance was estimated as only 0.2\%. 
 
The electromagnetic observations could be explained best through a multi-zone model, with a central acceleration region embedded in an ultraviolet photosphere that powers a
synchrotron-emitting outflow. This outflow provides the target for PeV neutrino production. 
The authors conclude that --- provided the coincidence of the two neutrino events with TDEs to be genuine --- the observations suggest that TDEs with relativistic outflows contribute to the flux of cosmic neutrinos \cite{Stein:2020xhk}. Shortly later \cite{Yang:2021qic}, AT2019fdr, an exceptionally luminous TDE candidate, was found to be coincident with another high-energy neutrino. The probability of finding two such bright events by chance is just 0.034\%. A consistent multimessenger description of these two and a third TDE with possible neutrino emission was developed in \cite{Winter:2022fpf}.

Since TDEs are rather rare events, in a very recent paper \cite{Langis:2026dyj}, TDECat, a comprehensive optical TDE repository, was used to investigate the coincidence of TDEs with IceCube neutrino events. Despite the individual cases of the TDE--neutrino coincidences reported before, no statistical association between optical TDEs and high-energy neutrinos was found. 
The question of TDEs as sources of high-energy neutrinos therefore stays open. 

\vspace{2mm}

\subsubsection{Gamma-Ray Bursts (GRBs)}
\label{GRB-subsubsection}
Gamma-ray bursts are the brightest events since the Big Bang. They occur at distances on the scale of billions of light-years and outshine whole galaxies. In order to explain the enormous observed energy release, one has to assume that the energy is not emitted isotropically but in intense jets that are pointing into our direction. They are assumed to be caused by the collapse of massive stars into black holes or the merger of two neutron stars.  Needless to say, they have been a top candidate for neutrino emission from the very beginning. 

GRB--neutrino correlations have been searched for by all underwater/under-ice telescopes: AMANDA \cite{IceCube:2007tsw},  NT-200 in Lake Baikal  \cite{Baikal:2009evy}, ANTARES \cite{ANTARES:2020vzs} and IceCube (starting with data taken with the half-completed detector \cite{IceCube:2011vle}). The most sensitive search has been performed by IceCube. In \cite{Abbasi:2022whi}, not only neutrinos from the prompt phase (milliseconds to minutes) were addressed but also regions of 14 days before and after the prompt phase.
For the period May 2011--October 2018, more than a thousand GRBs were selected and compared with muon-neutrino events recorded in IceCube. No evidence of a correlation between neutrino events and
GRBs was found. Prompt neutrino emission
from GRBs could be limited to 1\% and emission on timescales up to $10^4$\,seconds  to 24\% of the total diffuse flux observed by IceCube.

\begin{figure}[H]

\includegraphics[scale=0.7]{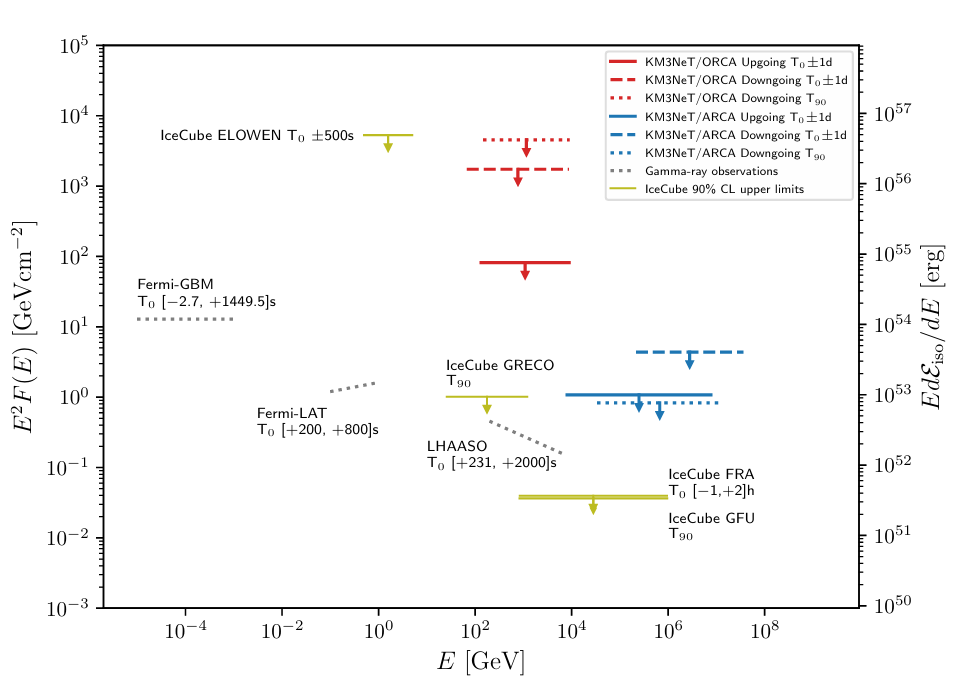}
\caption{The
90\% CL upper limits on $E^2 F(E)$, the energy-scaled time-integrated per-flavor neutrino
flux from GRB221009A for KM3NeT/ORCA, KM3NeT/ARCA 
and IceCube. Only the results for the $T_0 \pm 1$ day 
and $T_{90}$ searches are included, assuming a neutrino spectral index $\gamma$\,=\,2 ($T_{90}$ is the time window covering 90\% of the electromagnetic signal). For visualization purposes, the gamma-ray observations are also shown:
Fermi-GBM, Fermi-LAT, and LHAASO. The right axis indicates the differential isotropic equivalent energy. Figure taken from \cite{KM3NeT:2024nwb}. IceCube data published in~\cite{IceCube:2023rhf}.
}
\label{GRB221009Afig}      
\end{figure}

The strongest GRB ever observed was GRB 221009A.
LHAASO detected gamma rays from 200 GeV up to
more than 10 TeV, the highest energy observed so far from a GRB. Fermi-LAT registered gamma rays with energies close to 100 GeV.  
A GRB as bright as GRB221009A is expected to occur only once every ten thousand years.
But, even for this spectacular GRB, no excess of neutrino events has been observed by any of the neutrino telescopes. Figure \ref{GRB221009Afig} (taken from \cite{KM3NeT:2024nwb}) shows the neutrino limits obtained by the still rudimentary ARCA and ORCA detectors (10\% of the full configurations planned) and IceCube. They are compared to the results obtained by FERMI and LHAASO. The strong IceCube
upper limits \cite{IceCube:2023rhf} significantly constrain existing neutrino emission models from that source; see \cite{IceCube:2023rhf} for a detailed discussion.

\subsubsection{Gravitational Waves from Collapses}
\label{GW-subsubsection}

Gravitational waves are produced in a number of cosmic events, the strongest being cataclysmic events such as colliding black holes (BHs) or colliding neutron stars (NSs). If the primary objects are so massive that they collapse into a black hole almost immediately upon contact (BH/BH, BH/NS and part of the NS/NS mergers), no accretion disk or intermediate hypermassive neutron star will be left, and, consequently, no (or almost no) electromagnetic or neutrino signals emerge.  For NS/NS mergers, however, multimessenger studies including high-energy electromagnetic radiation and neutrinos have a chance to provide information on the origin of the high-energy emission and the dynamics and structure of the underlying processes.

\begin{figure}

\includegraphics[scale=1.1]{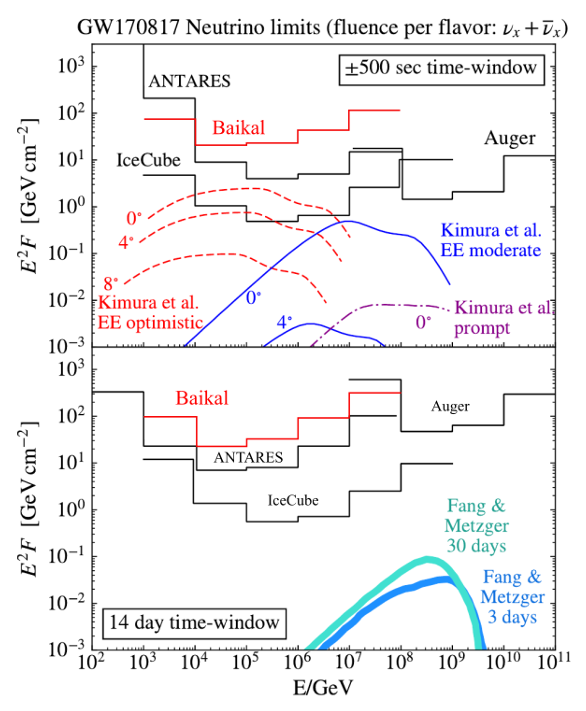}
\caption{ 
The 90\% CL upper 
 limits on
the neutrino spectral fluence from GW\,170817 during
a $\pm$500\,s window centered on the GW trigger time
(\textbf{top }panel) and a 14-day window following the GW
trigger (\textbf{bottom} panel). For each experiment, limits are
calculated separately for each energy decade assuming
a spectral fluence $F(E) = F_{up} \times (E/\text{GeV})^{-2}$ in that
decade only. Also shown are predictions by neutrino
emission models (see \cite{ANTARES:2017bia} for details).
}
\label{NSNS}      
\end{figure}

Since the discovery of gravitational waves in 2015 \cite{LIGOScientific:2016aoc}, a multitude of searches have been undertaken to detect accompanying neutrinos, starting with a follow-up of the LIGO Event GW\,150914 by ANTARES and IceCube \cite{ANTARES:2016qdk} --- without any positive signal.  In 2017, the Advanced LIGO and Advanced Virgo observatories recorded, for the first time,  gravitational waves from a binary neutron star (NS/NS) merger. The Fermi Gamma-Ray Burst Monitor (Fermi-GBM) and the Anticoincidence Shield for INTEGRAL 
recorded a short gamma-ray burst following the event. The electromagnetic signal not only indicated particle acceleration by the source but also allowed a precise location of the event. The ANTARES, IceCube, and Pierre Auger Observatories took the precise coordinates to search for high-energy neutrinos from the merger in the GeV\,--\,EeV energy range \cite{ANTARES:2017bia}. No neutrinos directionally coincident with the
source were detected within $\pm500$ s around the merger time. Also, no MeV neutrino burst signal coincident with the merger was detected (see Section \ref{SN-subsection} for the method to detect MeV neutrino bursts). Extending the search to a
14-day period following the merger also did not provide evidence of neutrino emission. 
Shortly later, the Baikal-GVD collaboration (at that time only running two of the present 16 clusters) published their limits \cite{Baikal-GVD:2018cya}.  Figure \ref{NSNS}, taken from \cite{Baikal-GVD:2018cya},
shows the IceCube, ANTARES, Auger  and Baikal-GVD limits of the neutrino flux from GW 170817 and compares them to model calculations. 
The non-detection of this gravitational wave event is consistent with model predictions of short GRBs observed at a large off-axis angle.

\subsection{Supernovae}
\label{SN-subsection}

The core-collapse neutrinos from SN 1987 detected almost forty years ago had energies in the MeV range. Although designed for non-thermal neutrinos with energies larger than 100 GeV, large neutrino telescopes are also able to detect neutrino bursts in the MeV range provided that the noise rates induced by the ambient medium are low. This condition applies in particular to IceCube. Due to the low temperatures and the absence of light from $K^{40}$ decays and biomatter, the photomultiplier dark noise rates are extremely low. Therefore, IceCube would be able to identify the collective rise in all the photomultiplier rates induced by a burst of MeV neutrinos. 
A supernova at the galactic edge (30 kpc) would be detectable with a significance of 
$20\,\sigma$ and a supernova in the Large Magellanic Cloud (50 kpc) with $6\,\sigma$, respectively \cite{IceCube:2011cwc}.

Note that just noise rates and not reconstructed events with their energy and direction would be observed, limiting the degree of information compared to detectors like Super-K or JUNO.  On the other hand, no other existing detector could measure the time profile as accurately as IceCube. 

IceCube is a member of SNEWS (Supernova Early Warning System) \cite{Scholberg:2008fa}. The SNEWS network would help optical telescopes to observe the SN signal from the very beginning since core-collapse neutrinos arrive hours before the optical onset of the signal. 


The possibility to detect MeV neutrinos from SN bursts has also been explored by the KM3NeT collaboration \cite{KM3NeT:2021moe}.
A supernova burst at 30 kpc distance could be detected with $7\,\sigma$ significance for a mass of the progenitor star of 27 solar masses, but the Large Magellanic Cloud would not be in reach. KM3NeT's smaller significance compared to IceCube 
($20\,\sigma$ for a star with 20 solar masses at 30 kpc distance) is due to the
much higher noise rates of the photomultipliers, which are mostly related to $K^{40}$ decays.

\begin{figure}[H]

\includegraphics[scale=0.7]{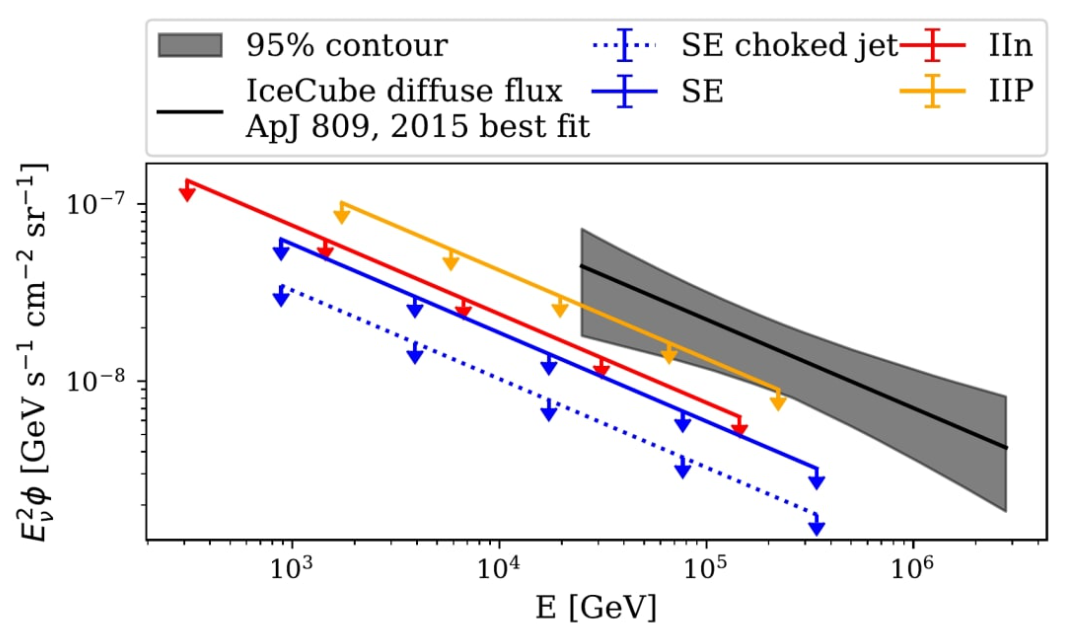}
\caption{ 
The 90\% CL 
 upper limits on the contribution of different
SN types to the diffuse neutrino flux assuming
an $E^{-2.5}$ energy spectrum compared to the measured diffuse astrophysical neutrino flux (gray band). 
The energy range plotted is the
central 90\% energy range of the analyzed neutrino sample.
Figure taken from \cite{IceCube:2023esf}.
}
\label{SN-HE}      
\end{figure}

The {\it remnants} of supernova explosions are also supposed to be sources of neutrinos, although in the GeV-TeV range rather than in the MeV range. Young supernova remnants like RX J1713.7-3946 and RX J0852.0-4622 (Vela Jr) belong to the strongest galactic gamma-ray sources and therefore are also candidates for GeV to TeV neutrino emission. With the advent of cubic-kilometer detectors the detection probability for neutrinos from these sources increased (see, e.g., \cite{Costantini:2004ap}). A search for a possible correlation between seven years of IceCube neutrino data and a catalog containing more than 1000 core-collapse supernovae in nearby galaxies \cite{IceCube:2023esf} provided upper limits only. Supernovae of types IIn, IIP and stripped-envelope SNe contribute less than 34\%, 60\% and 27\%, respectively, to the diffuse neutrino flux; see Figure\,\ref{SN-HE}.

A next galactic supernova would be a spectacular event, not only for the burst neutrinos but also for non-thermal neutrinos generated by the supernova remnant in the year after the collapse.
In \cite{Murase:2017pfe} the number of neutrino-generated throughgoing muon tracks in a cubic-kilometer detector from a supernova remnant at 10 kpc distance is estimated to be up to $10^3$ from an SN-type II-P (like SN 1987A) and $3 \times 10^5$ from an SN-type IIn.

\vspace{0.5cm}

\section{Synergies}
\label{Synergies-section}

\subsection{Sky Coverage}
\label{coverage-subsection}

Thirteen years ago, IceCube discovered a diffuse flux of high-energy cosmic neutrinos and opened a new window to the universe. At that time, the telescopes in the Mediterranean Sea (ANTARES) and Lake Baikal (NT200), with geometrical volumes two orders smaller than IceCube, could add anything beyond the IceCube results only in regions of the sky that were not covered by IceCube. And even this statement must be put in perspective since, with starting-track analyses or contained cascades, the full sky is accessible for IceCube. 

Meanwhile, with KM3NeT and Baikal-GVD, IceCube is going to be joined by partners of similar power.
It is clear that the diversity of tools and locations necessitates networking of neutrino telescopes. Beyond that, the multimessenger character of the sources necessitates multimessenger methods---a wide field of synergies!

\begin{figure}

\includegraphics[scale=0.5]{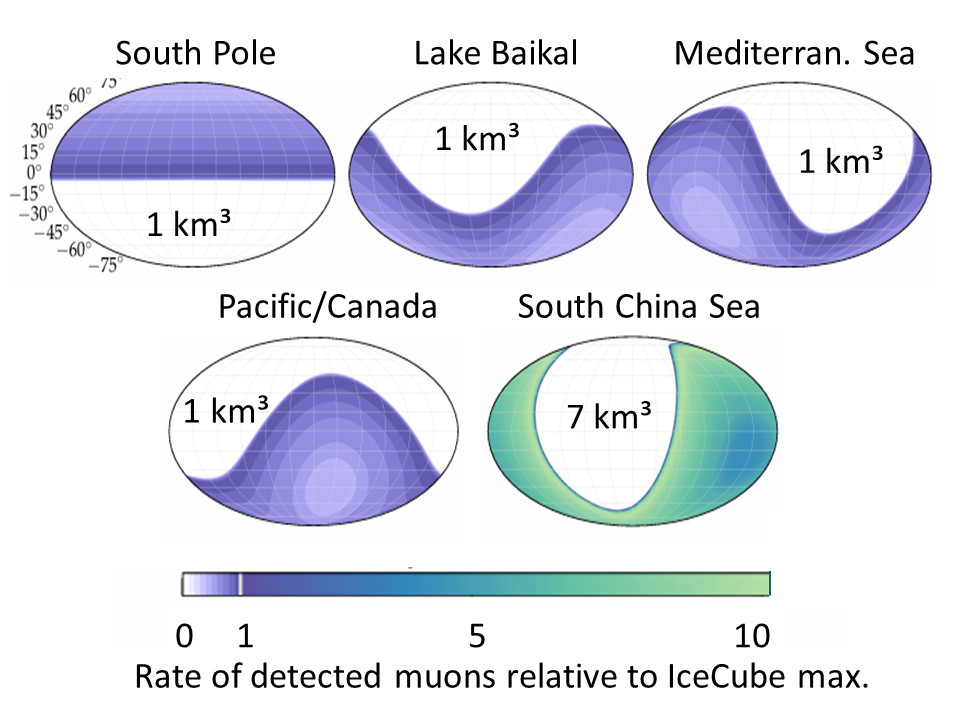}
\caption{ 
Expected rate 
 of muon tracks detected in present and next-generation high-energy optical Cherenkov neutrino telescopes. For each detector, the expected event rate is
computed and integrated over reconstructed muon energy
above 100 GeV and zenith angles up to $5^{\circ}$ above the horizon. Event rates are computed for an $E^{-2}$
neutrino energy spectrum and an instantaneous date and time (1 January 2025-00:00 UTC)
and are expressed relative to the maximum event rate achievable in IceCube. 
Figure modified from \cite{Schumacher:2025qca}.
}
\label{Sky-Coverage}      
\end{figure}

Continuous monitoring of the full sky by neutrino detectors at different locations has been investigated in much detail in \cite{Schumacher:2021hhm,Schumacher:2025qca}. Figure \ref{Sky-Coverage} demonstrates the instantaneous sky coverage (here: 1 January 2025-00:00 UTC)
of five detectors at different locations: the South Pole, Mediterranean Sea, Pacific Coast (Canada), Lake Baikal and South China Sea. It is obvious that the sky coverage by these five detectors is almost 100\%; i.e., at any time, each spot on the sky can be observed via track-like events by at least one detector. 
For steady sources, most of the sky will sweep across the field of view of three to four detectors.

\begin{figure}

\includegraphics[scale=0.6]{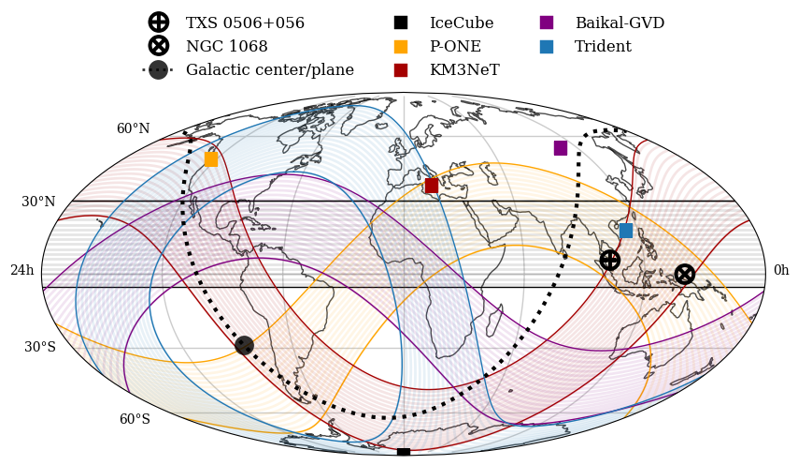}
\caption{Instantaneous 
 regions of highest detection efficiency for
various neutrino telescopes. The colored regions cover a zenith region 
between $5^{\circ}$ above and $30^{\circ}$ below the horizon
(transmission of Earth at $30^{\circ}$ below the horizon is about 80\% for 100 TeV and
40\% for 1 PeV). 
As in the previous figure, this is an instantaneous snapshot; the bands move as Earth rotates. Figure courtesy Lisa Schumacher.
}
\label{Sky-Plenum}      
\end{figure}

The situation looks different if one considers only the regions where extremely energetic neutrinos are not absorbed by Earth and exceed the contribution from atmospheric neutrinos, i.e., the regions where the discovery potential is highest. Figure \ref{Sky-Plenum} addresses the same neutrino telescopes as in Figure\,\ref{Sky-Coverage}. Here, however, the colored regions cover a zenith region between $5^{\circ}$ above and $30^{\circ}$ below the horizon.
The edges of the bands correspond to a discovery potential of half its maximum value. But, even for this limited zenith angle range, almost 80\% of the sky is in the field of view of at least one of the detectors at any~time.

\subsection{Combined Skymaps}
\label{combined-subsection}

Combining skymaps has been practiced by IceCube and ANTARES for a decade; see \cite{ANTARES:2015moa,ANTARES:2020srt,ANTARES:2020leh}. For the coming decade, the sensitivity for {\it steady} sources in the Northern Hemisphere will likely be dominated by IceCube due to the long data taking (in 2035: 25~years compared to 5--7 years with the full two-cluster KM3NeT configuration). For sources in the Southern Hemisphere, however, the benefit of combined skymaps will set in much~earlier. 

The potential contribution of Baikal-GVD is hard to estimate. The present results (see Figure~\ref{B-and-A}) are based on tracks triggered and reconstructed only within single clusters. The reconstruction of tracks using more than just single GVD clusters appears to be mandatory to increase the present statistics of only $\sim$$10^3$ events from three years of data taking. The time lag for P-ONE and the Chinese detectors, not to mention IceCube-Gen2, will be much larger than for KM3NeT. On the other hand, only multi-cubic kilometer detectors like those will make the big step towards skymaps populated by dozens of sources and not just a few.

For {\it transient} sources it is only the actual detector size and its angular resolution that counts. Therefore the benefit of combined analyses sets in much earlier than for steady sources, where newcomers have to catch up with older detectors regarding lifetime.

Combining skymaps requires the knowledge of the declination-dependent sensitivity of participating experiments (typically given for a certain value of the spectral index of the energy spectrum, e.g., $-$2 or $-$2.5) and the  signal and background probability density functions (PDFs). Such projects can be performed on a single-case basis on the basis of MoUs between two experiments or in the future as part of the Global Neutrino Network (see Section \ref{GNN-subsection}).

\subsection{Alerts, Follow-Ups and Fast Responses}
\label{alerts-subsection}


{\it Alert messages} 
 to the astronomical community are based on single exceptionally energetic neutrino candidates with a high likelihood to be of cosmic origin. In addition, sub-threshold data can be sent to a central server, combined with data from other instruments, and possibly also result in an alert. Examples are the Supernova Early Network, SNEWS~\cite{Scholberg:2008fa}, or the Astrophysical Multimessenger Observatory Network, AMON \cite{AyalaSolares:2019iiy}.


{\it Follow-up} messages 
are based on events coming from close-by sky localizations within a certain time window (''multiplets''). For their identification, the full skymap is checked for local clusters building up over some time. In case a ``cluster criterion'' (a sufficiently large number of events accumulated over a sufficiently small time interval) is fulfilled, a message is sent to gamma-ray telescopes (Gamma-Ray Follow-Up, GFU) and/or optical astronomers. 
Care has to be taken in both cases that the false alarm rate (FAR) stays below a threshold that is acceptable to the follow-up instruments. 

{\it Fast response analysis} (FRA) searches for neutrino emission coincident with recent astrophysical transients like gravitational waves, GRBs, or AGN flares. 

\begin{figure}[H]

\includegraphics[scale=0.75]{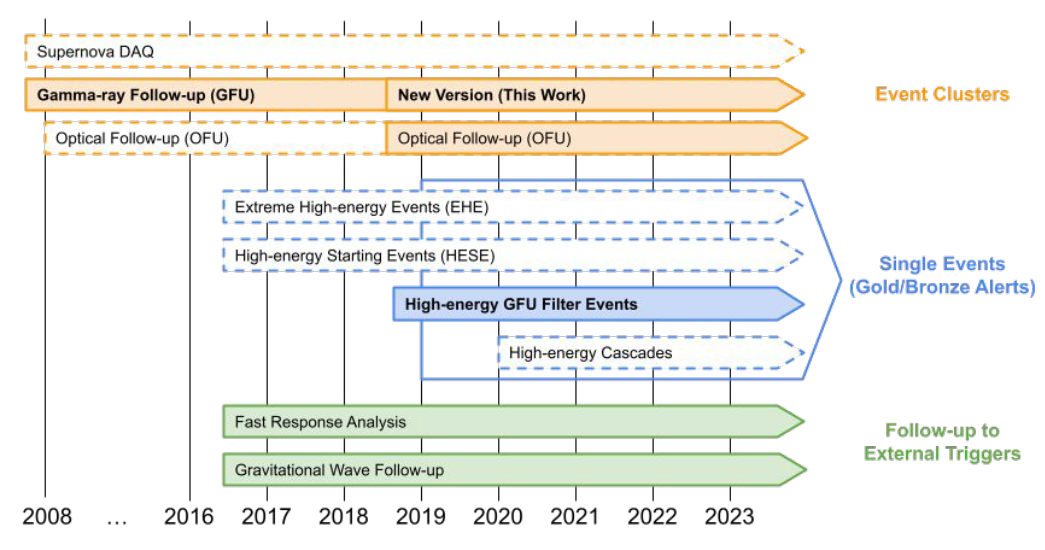}
\caption{IceCube: Overview on alerts, follow-ups and fast response programs.
Figure taken from a talk of Sarah Mancina at NuTel 2023
(\url{https://agenda.infn.it/event/33107/contributions/206336/}). 
}
\label{alerts}      
\end{figure}

The alert and follow-up systems of neutrino telescopes are described, e.g., in \cite{ANTARES:2011ybh,ANTARES:2022gyy} (ANTARES), \cite{Cecchini:2025wlu} (KM3NeT), \cite{IceCube:2016cqr,IceCube:2020mzw,VERITAS:2021mjg}  (IceCube) and \cite{Allakhverdyan:2025krm} (Baikal-GVD). Figure \ref{alerts} gives an overview of IceCube's alert, follow-up and fast response programs.

IceCube's gold and bronze alerts (about 30 per year) are publicly broadcasted through
AMON and also sent to private partners (defined via Memoranda of Understanding, MoU), the same applied to ANTARES (which, apart from a ``high-energy selection'' with $\sim$25 alerts per year, also had a selection ``very close to local galaxies'' with $\sim$10 alerts per year). Baikal-GVD registers about one high-energy cascade per cluster and year and releases some fraction of those as Astronomer's Telegram (ATel). Receivers are (either directly or via AMON) robotic optical telescopes (like MASTER), X-ray satellites (SWIFT and INTEGRAL), radio telescopes and TeV gamma-ray telescopes.
The typical observation efficiencies are 50--70\% (X-ray/optical) and 20\% (radio). 
Follow-ups of gamma-ray telescopes are typically regulated via MoUs and have a much higher threshold to be released.

As a reminiscence I remind readers that the foundations for follow-up programs were laid about 20 years ago. In 2005, the AMANDA collaboration reported the detection of two neutrinos from the direction of the blazar 1ES 1959+650. They coincided with TeV gamma-ray flares observed by the Whipple Telescope in May and June 2002, respectively. The TeV flares were ``orphan flares''; i.e., they were not accompanied by X-ray flares.
The probability of a chance coincidence between background events and the 1ES 1959+650 orphan flare was rather small ($\sim$$10^{-3}$). However, the two events were not detected in a blind analysis, so the statistical significance of the coincidence could not be estimated. Also, to explain the two events, an unreasonably high neutrino flux had to be assumed. It was this event that led to AMANDA's Gamma-Ray Follow-Up program~\cite{Franke:2011xjk}, 
motivating gamma-ray astronomers to point their telescopes in the direction of timely clustered events, thereby increasing the probability of simultaneous observations of gamma-ray flares and neutrinos.

\subsection{The Global Neutrino Network}
\label{GNN-subsection}

The Global Neutrino Network (GNN; see  \url{https://www.globalneutrinonetwork.org/}) was established in 2013 as an umbrella organization of the neutrino telescope projects ANTARES, Baikal-GVD, IceCube and KM3NeT. Meanwhile two new projects have joined GNN: P-ONE and RNO-G (see Section \ref{NTs}).

GNN provides a forum for cross-coordination between the individual collaborations and the development of common strategic objectives. The goals of the GNN include the coordination of alert and multimessenger policies, exchange and mutual checks of software, creation of a common software pool, developing standards for data representation and cross-checks of results with different systematics. It obviously does not make much sense to invent tools and technologies twice if they have already been developed elsewhere. Increasing the sensitivity by combining data from different experiments and cooperation in multimessenger campaigns and alert strategies are among the top goals of the GNN. 

The GNN community hosts two biannual meetings, the internal GNN Meeting and the Symposium VLV$\nu$T (Very Large Volume Neutrino Telescopes). Since 2016, the GNN Community has also authored the  newsletter “GNN Monthly”. It is released as a publicly available document since November 2020 (see the GNN website). 

Several working groups have been formed or are likely to be formed under the umbrella of the GNN, e.g., a working group on data formats and another one on AI/ML applications. Machine learning (ML) procedures are going to outperform likelihood-based methods, a promising field to share tools. Combined IceCube and ANTARES data have already been used in searches for sources of astrophysical neutrinos or in dark matter searches. With KM3NeT taking shape, the combination of data in full-sky source searches --- so far more of a practice exercise than a truly promising project --- will become a more prominent tool.

\section{Conclusions}
\label{Conclusions-section}

Neutrino astronomy has opened a new window to the universe. The energy density of energetic cosmic neutrinos is comparable to that of high-energy cosmic rays, underscoring the importance of neutrinos for understanding cosmic rays and discovering their sources. However, with the exception of the Milky Way and two point-like sources, the neutrino sky still appears as a diffuse glow. Before the term ``astronomy'' can be used with a similar justification to that used by gamma-ray astronomers, the celestial landscape of high-energy neutrinos has to be populated by many more sources, with a much better understanding of their properties, than presently.

There is no shortage of hints of more than the two mentioned point-like sources. While blazars {\it alone} cannot explain the diffuse flux measured by IceCube, there are strong indications that they significantly contribute. The same applies to Seyfert galaxies. 
There are other source classes with a higher cosmic density but a smaller effective neutrino flux than blazars or Seyfert galaxies---e.g., starburst galaxies. They may also significantly contribute to the diffuse flux but are difficult to identify.

Excellent angular resolution is a must to identify point sources. As important as the {\it cascade channel} has turned out to be for diffuse and quasi-diffuse sources like the Galactic Plane, it will be only the {\it track channel} that can provide the signal-to-background ratio that is necessary to identify steady point sources. 

It is obvious that one needs more and larger neutrino telescopes. Cubic-kilometer neutrino telescopes in the Northern Hemisphere will be a first step, in particular regarding sources in our own galaxy. But, for clear identification of all those sources that presently barely rise above the surface of the diffuse neutrino sea (the ``2--3\,$\sigma$ class''), the ten-kilometer scale seems to be inevitable. A network of such multicubic-kilometer detectors will presumably need another decade to become reality.
It is, however, not the size alone that matters. Rather than only building larger and better detectors, the close cooperation of all neutrino telescopes is of key importance to make astronomy a thriving scientific field.

A significant role in this success has been played by the multimessenger approach. Actually, catalog searches and alert messages have provided the most intriguing indications or evidence for point sources; see Section \ref{point-source-section}. With larger and more neutrino telescopes, this aspect will become even more important. A neutrino signal alone can certainly prove the hadronic nature of a source but will never provide the full picture: it calls for interpretation in the context of electromagnetic signals. Multimessenger analyses, as well as coordinated analyses among neutrino telescopes, combined skymaps and common alert strategies, as described in Section \ref{Synergies-section}, will be key ingredients for the success of neutrino astronomy.

\vspace{6pt}


\acknowledgments{I thank Markus Ackermann and Jakob van Santen for helpful comments and corrections. I also thank the anonymous referees for their valuable feedback and corrections.}

\reftitle{References}


\end{document}